\documentclass[manuscript]{aastex}

\shorttitle{Proto-Brown Dwarf candidates in Serpens}
\shortauthors{Riaz et al.}

\begin{document}

\title{A multi-wavelength characterization of proto-brown dwarf candidates in Serpens}

\author{B. Riaz}
\affil{Max-Planck-Institut f\"{u}r Extraterrestrische Physik, Giessenbachstrasse 1, 85748 Garching, Germany}
\email{briaz@mpe.mpg.de}

\author{E. Vorobyov}
\affil{Institute of Astrophysics, University of Vienna, Vienna 1180, Austria; Research Institute of Physics, Southern Federal University, Rostov-on-Don 344090, Russia}

\author{D. Harsono}
\affil{Universität Heidelberg, Zentrum für Astronomie, Institut f\"{u}r Theoretische Astrophysik, Albert-Ueberle-Str. 2, 69120, Heidelberg, Germany}

\author{P. Caselli}
\affil{Max-Planck-Institut f\"{u}r Extraterrestrische Physik, Giessenbachstrasse 1, 85748 Garching, Germany}

\author{K. Tikare}
\affil{IRAP, BP 44346 - 31028 Toulouse Cedex 4, France}

\author{O. Gonzalez-Martin}
\affil{Instituto de Radioastronomía y Astrofísica (IRyA), UNAM, Antigua Carretera a P\'{a}tzcuaro \# 8701, Col. Ex Hacienda San Jos\'{e} de la Huerta, Morelia, Michoac\'{a}n, M\'{e}xico, C.P. 58089}

\begin{abstract}

We present results from a deep sub-millimeter survey in the Serpens Main and Serpens/G3-G6 clusters, conducted with the Submillimetre Common-User Bolometer Array (SCUBA-2) at the James Clerk Maxwell Telescope. We have combined {\it Herschel} PACS far-infrared photometry, sub-millimeter continuum and molecular gas line observations, with the aim to conduct a detailed multi-wavelength characterization of `proto-brown dwarf' candidates in Serpens. We have performed continuum and line radiative transfer modeling, and have considered various classification schemes to understand the structure and the evolutionary stage of the system. We have identified four proto-brown dwarf candidates, of which the lowest luminosity source has an $L_{\rm bol} \sim$0.05 $L_{\sun}$. Two of these candidates show characteristics consistent with Stage 0/I systems, while the other two are Stage I-T/Class Flat systems with tenuous envelopes. Our work has also revealed a $\sim$20\% fraction of mis-identified Class 0/I/Flat sources that show characteristics consistent with Class II edge-on disk systems. We have set constraints on the mass of the central object using the measured bolometric luminosities and numerical simulations of stellar evolution. Considering the available gas+dust mass reservoir and the current mass of the central source, three of these candidates are likely to evolve into brown dwarfs. 



\end{abstract}


\section{Introduction}
\label{intro}

The `Serpens Main' or the `Serpens Core' cluster has been the subject of several studies at various wavelength regimes to understand the influence of the cluster environment in the process of star and planet formation. This young embedded cluster is concentrated into the north-western (NW) and south-eastern (SE) clumps or sub-clusters (Casali et al. 1993; Davis et al. 1999). About 45$\arcmin$ south of the Serpens Main lies another active star-forming region, `Serpens/G3-G6' (Serp/G3-G6), which is a complex around four T Tauri stars named G3 to G6 in a 30$\arcsec$ field (e.g., Harvey et al. 2006; Djupvik et al. 2006). 

Several infrared surveys conducted in these clusters have identified more than 200 young stellar objects (YSOs) in various evolutionary stages, from pre-stellar cores to Class 0 to the more evolved Class II young disk sources (e.g., Evans et al. 2003; Harvey et al. 2006; 2007ab; Evans et al. 2009). A high population of Class 0/I systems has been found in the NW and SE sub-clusters, while the more evolved Class II/III sources are dispersed over the larger region (e.g., Kaas et al. 2004; Harvey et al. 2006; 2007; Winston et al. 2007). Far-infrared {\it Herschel} observations have revealed new protostars (Bontemps et al. 2010; Konyves et al. 2010). In the sub-millimeter/millimeter regime, Enoch et al. (2007; 2009) identified 12 continuum sources in CSO Bolocam observations at 1 mm. A more extensive sub-millimeter catalogue in Serpens Main has been constructed using the SCUBA bolometer array camera on the James Clerk Maxwell Telescope (JCMT), named the SCUBA Legacy Catalogue (SLC; Di Francesco et al. 2008). Four pre-stellar cores were later identified in the SLC (Sadavoy et al. 2010). Recent observations with CARMA have identified a network of filament structures that may fragment into protostellar cores (Lee et al. 2015).

We have conducted a deep sub-millimeter continuum survey of the Serpens Main and Serp/G3-G6 clusters, using the Submillimeter Common-User Bolometer Array (SCUBA-2; Holland et al. 2013) at the JCMT telescope. We have combined near-, mid-, and far-infrared observations with sub-millimeter continuum and molecular gas line observations, with an aim to conduct a detailed multi-wavelength characterization of `proto-brown dwarf' candidates in Serpens. Previous SCUBA Legacy observations conducted in these clusters reached a (median) rms of $\sim$70 mJy/beam at 850\,$\mu$m (Di Francesco et al. 2008). Our survey reaches an rms that is more than $\sim$10 times deeper than the SLC, thus enabling the detection of YSOs at the very low-mass end that remained undetected in previous sub-millimeter observations. Our multi-wavelength characterization has also led to the identification of YSOs that may have previously been `mis-identified' as an early-stage Class 0/I embedded source, based on their near- to mid-infrared spectral index, and show characteristics that are more consistent with Class II objects with edge-on disks.

The distance to Serpens is still a matter of debate. Several studies have adopted a distance of 260$\pm$10 pc from Strai\u{z}ys et al. (2003), based on the spectral and luminosity class of the observed stars. The 260 pc estimate is probably correct to the front of the cloud or to the front edge of the Aquila rift, but the depth of the cluster behind it is unknown. A common distance of 260 pc has been suggested for the Aquila rift and Serpens Main cluster, based on the argument that these are parts of the same star-forming region (Bontemps et al. 2010). A farther distance of 415 pc has been obtained by Dzib et al. (2010) using the VLBA parallax measurement of the young AeBe star EC95 in the SE sub-cluster of Serpens Main. Winston et al. (2010) obtained a distance of 360 pc for the Serpens core, based on the X-ray luminosity function of the cluster. Their revised estimate based on new X-ray data is 260 pc for Serpens South and closer to 310 pc for Serpens North (E. Winston, {\it priv. comm.}). Another test of the distance comes from placing the spectroscopically confirmed members on the HR diagrams, and a distance of 415 pc does not correlate with the age of the cluster (Winston et al. {\it in prep.}). Our sample selection criteria and some of the analysis is based on the luminosity estimates from Evans et al. (2009), who had adopted a distance of 260$\pm$10 pc for Serpens. We have therefore opted for a 260 pc distance as it is more suitable for this work, rather than the recent estimate of 415 pc.

The sample and various observations are described in Sect.~\ref{obs}. Results from the continuum and line radiative transfer modeling, along with a description of the bonafide detections, and the mis-identified objects are presented in Sect.~\ref{results}. A discussion on the classification of the YSOs using different schemes, estimates on the total mass of the system, and the likelihood of the proto-brown dwarf candidates evolving into a sub-stellar object is presented in Sect.~\ref{discussion}.

\section{Sample, Observations and Data Reduction}
\label{obs}

\subsection{Sample}
\label{sample}

The most detailed hunt for YSOs in the Serpens Main and Serp/G3-G6 clusters was conducted under the {\it Spitzer} ``Cores to Disks'' (c2d) legacy project (Evans et al. 2003). We searched the c2d catalog for all early-stage (Class 0/I/Flat) YSOs with bolometric luminosity, $L_{\rm bol}$ $\leq$ 0.3 $L_{\sun}$. We selected this limit since it corresponds to the typical boundary between very low-mass (VLM) and low-mass objects (e.g., Chabrier \& Baraffe 1997; 2000). Considering the mass-luminosity relations based on the non-accretion models by Baraffe et al. (2003), an $L_{\rm bol}$ $\sim$ 0.3 $L_{\sun}$ would correspond to a stellar mass of $\sim$0.2-0.3 $M_{\sun}$, assuming an age of $\sim$1 Myr for the cluster. Based on the same models, sub-stellar mass objects ($M_{*}$$<$0.08 $M_{\sun}$) or brown dwarfs (BDs) at this age would have $L_{\rm bol} \leq$ $\sim$0.05 $L_{\sun}$. Our target list consists of 28 VLM/BD objects, 13 of which are classified as Class 0/I and 15 as Class Flat sources, as listed in the {\it Spizer} c2d catalog. 

It is important to note a possible uncertainty in converting $L_{\rm bol}$ into stellar mass using non-accreting stellar evolutionary models, as in Baraffe et al. (2003), which do not include the contribution to the luminosity from ongoing accretion. The bolometric luminosity in such a case can only provide an upper limit to the stellar luminosity or mass of the source. We have taken into account the accretion effects and set constraints on the mass of the central object using new, more accurate accreting models, as presented in Sect.~\ref{totalmass}.

\subsection{Sub-millimeter Continuum Observations}
\label{submm-obs}

\begin{figure}
  \centering              
     \includegraphics[width=5in]{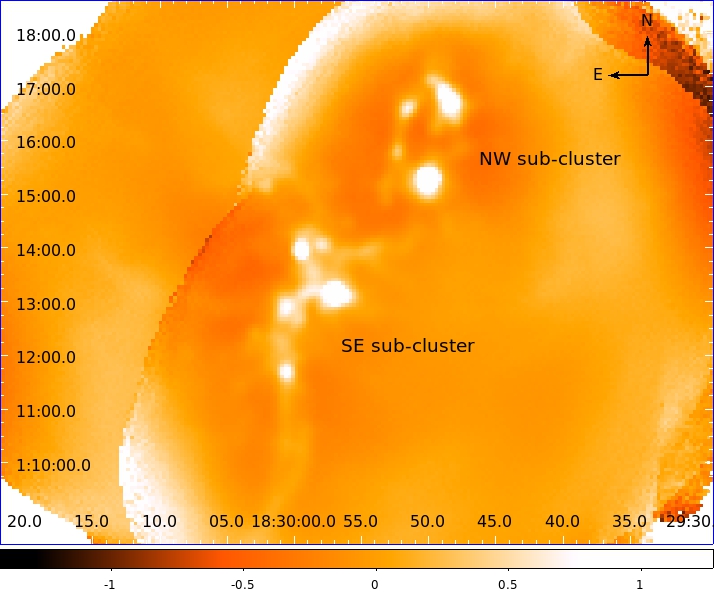} \\
     \vspace{0.2in}
     \includegraphics[width=5in]{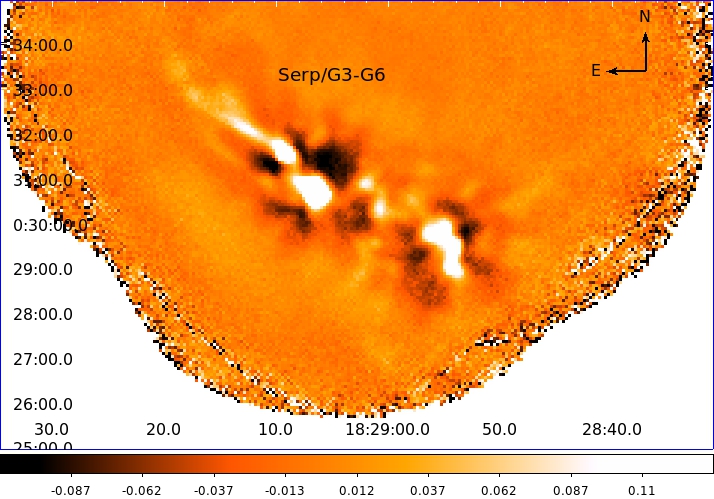}    
     \caption{The SCUBA-2 850 $\mu$m maps in Serpens Main ({\it top}) and the Serp/G3-G6 ({\it bottom}) clusters. The color scale at the bottom shows the intensity in units of Jy beam$^{-1}$. North is up, east is to the left. }
     \label{maps}
  \end{figure}

We obtained sub-millimeter continuum observations using the JCMT SCUBA-2 (Holland et~al. 2013) instrument. SCUBA-2 provides dual-wavelength observations at 450 and 850$\mu$m. The default map pixels are 2$\arcsec$ and 4$\arcsec$ and the half-power beam width is 7.5$\arcsec$ and 14.5$\arcsec$ in the 450 and 850\,$\mu$m bands, respectively. The observations were obtained in the months of April-June, 2014 (PID: M14AU05) in Grade 2 weather (225 GHz opacity of 0.06). We used the CV Daisy observing mode, and also applied a matched-beam filter which utilises the full flux in the beam. Using this setup, we obtained a 1-$\sigma$ {\it rms} of $\sim$3\,mJy/beam at 850\,$\mu$m. The reduction and calibration of the SCUBA-2 data was conducted using the Sub-Millimetre User Reduction Facility (SMURF), which is available under the Starlink package. The final science maps were produced using the various steps and the default configurations described in the map-making process DIMM (Dynamic Iterative Map-Maker; Chapin et~al. 2013). We used the KAPPA and PICARD softwares for the post-processing of the science maps. To improve the point-source detectability, the PICARD recipe SCUBA2\_MATCHED\_FILTER was applied to the flux-calibrated science maps. This recipe fits a single Gaussian point spread function (PSF), centered over every pixel in the map, and applies a background suppression filter to remove any residual large-scale noise. The half-power beam width in the respective bands was used as the full-width at half maximum of the Gaussian fit. The images are convolved with the modified PSFs. This technique is recommended for faint source extraction and has been utilized before in e.g. creating the SCUBA Legacy catalog (Di Francesco et al. 2008). The matched filtering technique is valuable to extract the optimum flux from such low-luminosity objects, and helps to minimize the contamination from the surrounding cloud material. The full scale of the 850 $\mu$m maps are shown in Fig.~\ref{maps}.




\subsection{Mid-infrared Spectroscopy}

We obtained mid-infrared low-resolution (R$\sim$600) N-band spectroscopy with the CanariCam spectrograph on the Gran Telescopio Canarias in June, 2015. The observations were conducted under conditions with PWV$\leq$8 mm. We observed using the chop-nod mode, and a slit width of 0.5$\arcsec$. We obtained a 1-$\sigma$ point source sensitivity of 55 mJy at 10$\mu$m. For data reduction, we used the RedCan pipeline developed by Gonzalez-Martin et al. (2013). The final products from this pipeline are flux-calibrated spectra combined from multiple exposures of each science target. We estimate a SNR$\sim$10 for the spectra; the SNR is lower at the very base of the spectrum near $\sim$9.5 $\mu$m. For three sources, low-resolution Spitzer/IRS spectra were available in the archives, and these were extracted and calibrated using the Spitzer IRS Custom Extraction (SPICE) software. Further details on the processing of the Spitzer/IRS spectra are provided in Riaz (2009). The mid-infrared spectra are shown in Fig.~\ref{MIRspec}.

\subsection{Archival Data}

\subsubsection{Herschel PACS Observations}

We have analyzed the archival Herschel PACS 70 $\mu$m, 100 $\mu$m, and 160 $\mu$m scans of Serpens Main and Serp/G3-G6 clusters. These observations were taken from the programmes ObsIDs 1342229080 and 1342206676. We analyzed the final pipeline-processed ``Level 2.5'' data, and used the rectangular-sky photometry task provided in the Herschel Interactive Processing Environment (HIPE). The rectangular photometry task applies aperture photometry to both a circular target aperture and a rectangular sky aperture. The sky intensity was estimated using the median-sky estimation algorithm. Since the sky intensity could vary with location, we selected four different background regions around the target and measured the sky-subtracted source flux in each sky region. The final source flux and error is the mean and the standard deviation of these four flux measurements.

\subsubsection{Molecular Line Observations}
\label{moldata}

We searched for molecular line observations in various archives, and were able to retrieve observations of the HCO$^{+}$ (3-2) transition obtained with the CSO heterodyne receiver (XFFTS spectrometer). The final processed CLASS files for the HCO$^{+}$ data were kindly sent by N. Evans and A. Heiderman, while some of the data was obtained from the CSO archives. The data reduction process of the HCO$^{+}$ data is discussed in Heiderman \& Evans (2015). There were no C$^{18}$O line observations available in the archives for the VLM/BD sources detected in the SCUBA- 2 maps. The molecular line spectra for the individual sources are discussed in Sect.~\ref{bonafide}.

\section{Results}
\label{results}

We have detected 7 Class 0/I and 8 Class Flat very low-luminosity YSOs in our SCUBA-2 sub-millimeter observations. A detailed description of the YSOs that have a bonafide detection in the 850\,$\mu$m band, with a signal-to-noise ratio SNR $\geq$ 5, is provided in Sect.~\ref{bonafide}. The bonafide detections are listed in Table~\ref{properties}. The PACS and SCUBA-2 photometry for these sources is listed in Table~\ref{phot}. Among the bonafide detections, there are four `mis-indentified' cases for which the evolutionary stage based on the physical characteristics appears inconsistent with their original SED class provided in the c2d catalog (Sect.~\ref{misclassified}). In addition, there are 6 proto-brown dwarf candidates that are undetected (SNR $<$ 2) in the SCUBA-2 maps, but show emission in the HCO$^{+}$ (3-2) molecular line (Sect.~\ref{pbds}). The discussion on sources with marginal sub-millimeter detection (SNR $\sim$2), and sources that are unresolved/confused in the SCUBA-2 and PACS maps due to close proximity to a more luminous protostar is presented in Sect.~\ref{appendix}.

\subsection{Radiative Transfer Modeling of the Spectral Energy Distributions}
\label{model}

Radiative transfer modeling of the spectral energy distributions (SEDs) can map the physical structure of the system, and is important to obtain estimates on the masses and sizes of the envelope and disk components, the envelope mass infall rate, the inclination angle of the system. It can also provide better constraints on the contribution to the total luminosity from the envelope and disk components. We have built the spectral energy distribution (SEDs) for the bonafide detections by compiling photometric data in the near-infrared from the UKIDSS and/or the 2MASS surveys, in the mid-infrared from the ALLWISE catalog, the PACS far-infrared photometry, and the SCUBA-2 sub-millimeter photometry. For 6 sources, the mid-infrared spectra have also been included in the fits. For three sources, the 1.1 mm photometry from CSO Bolocam observations (Enoch et al. 2009) is available. However, the 1.1 mm flux density is higher than the 850\,$\mu$m photometry for all three sources, due to which we have opted not to use the 1.1 mm point in fitting the SEDs. Any model that provides a good fit to the 1.1 mm point misses the far-infrared 70-160 $\mu$m and the sub-millimeter 450 and 850 $\mu$m points, and is a poor fit to the overall SED. The possible reasons for the discrepancy between 850 $\mu$m and 1.1 mm photometry are discussed in Sect.~\ref{caveats}. 

The SED modeling was conducted using the two-dimensional radiative transfer code by Whitney et~al. (2003). The main ingredients of the model are a rotationally flattened infalling envelope, bipolar cavities, and a flared accretion disk in hydrostatic equilibrium. For the circumstellar envelope, the angle-averaged density distribution varies roughly as $\rho \propto r^{-1/2}$ for $r \ll R_{c}$, and $\rho \propto r^{-3/2}$ for $r \gg R_{c}$. Here, $R_{c}$ is the centrifugal radius and is set equal to the disk outer radius. This includes only the current infalling part of the envelope. Beyond $r_{infall}$, the density decreases as $r^{-2}$ and most mass is out in that region for early (Stage 0) phase. The disk density is proportional to $\varpi^{-\alpha}$, where $\varpi$ is the radial coordinate in the disk midplane, and $\alpha$ is the radial density exponent. The disk scale height increases with radius, $h=h_{0}(\varpi / R_{*})^{\beta}$, where $h_{0}$ is the scale height at $R_{*}$ and $\beta$ is the flaring power. The disk extends from the dust sublimation radius, $R_{sub}$ = $R_{*} (T_{sub}/T_{*})^{-2.1}$, to the outer disk radius, $R_{disk,max}$. Here, $T_{sub}$ is the dust sublimation temperature which was set to 1600 K. The bipolar cavities in the models extend from the centre of the protostar to the envelope outer radius ($R_{max,env}$). We adopted the curved shaped cavity, which has the structure of $z = a\varpi^{\beta}$, where $\varpi = (x^{2} + y^{2})^{1/2}$ and the constant {\it a} is determined by a relation between the envelope radius and the cavity opening angle (Whitney et~al. 2003).

Table~\ref{SEDpars} lists the estimates for the various envelope and disk model parameters, based on the best model-fit (lowest $\chi^{2}$ value) to the observed SED. The uncertainties for each parameter represent the range in values obtained from the degeneracies in the top five model fits. The parameters that notably affect the model SEDs are the mass infall rate ($\dot{M}_{env}$), the inclination angle ($\theta_{in}$), and the centrifugal radius. An increase in $\dot{M}_{env}$ corresponds to a denser envelope, which implies that a lesser number of photons can escape through the cavity regions, thus producing lower near-infrared fluxes. The best model fits to the observed SEDs for the Class 0/I targets were obtained for $\dot{M}_{env}\sim$ 10$^{-5}$--10$^{-6}$ $M_{\sun}$ yr$^{-1}$ (Table~\ref{SEDpars}). Higher values for $\dot{M}_{env}$ results in a model with excess emission in the sub-millimetre regime, and does not fit the 850\,$\mu$m point. We can therefore constrain the $\dot{M}_{env}$ parameter using both the sub-millimetre and the near-infrared points. 

The amount of flaring in the system, particularly around $\sim$100$\mu$m, can be controlled with the centrifugal radius. For smaller values of $R_{c}$, the model SEDs show more flux at wavelengths $\geq$24 $\mu$m and lower fluxes shortward of $\sim$8 $\micron$. Decreasing $R_{c}$ implies that infalling material piles up closer to the central protostar. This results in fewer photons that can escape through the cavity walls and the disk upper layers, and shifts the far-infrared peak in the model SED towards shorter wavelengths. Likewise, the effects of the inclination angle on the model SED are also mainly seen in the far-infrared. A high-inclination system implies that there is more absorbing material in our line-of-sight. Thus with decreasing inclination angle, there is a decline in the far-infrared fluxes near 100 $\mu$m, and an increase in the near- and mid-infrared fluxes. The effects of increasing the outer radius of the envelope ($R_{\rm env,max}$) are mainly seen in the model SED at the longest wavelengths. This parameter was constrained by the far-infrared/sub-millimeter points, and a value of $\sim$1500-2000 AU provides a good fit. The depth, shape, and width of the 10$\mu$m silicate absorption feature provides additional constraints to the envelope and cavity parameters in the model fits. Increasing the envelope density results in a deeper silicate absorption profile, whereas increasing the cavity opening angle results in a shallower silicate feature. Among the disk parameters, the outer disk radius ($R_{\rm disk,max}$) was set to the same value as the centrifugal radius, while the inner disk radius is a few stellar radii ($\sim$3-5 $R_{sub}$). Results from the SED modeling for the individual sources are discussed in Sect.~\ref{bonafide}.

\subsection{Radiative Transfer Modeling of the Molecular Lines}

The HCO$^{+}$ (3-2) transition can probe the dense gas located in the inner regions of protostellar envelopes, and is known to be an important indicator of a Class 0/I source (e.g., van Kempen et al. 2009). The interpretation of HCO$^{+}$ data requires a simple model to verify the dynamics of the system. We adopt the `two-layer' model of Di Francesco et al (2001), based on Myers et al. (1996). The two layers are assumed to be a uniform slab layer characterized by a single temperature $T_{f}$ and $T_{r}$ for front and rear layer, respectively. The difference from the earlier Myers et al (1996) model is the addition of a continuum layer $T_{c}$ in between. The front layer is a representative of the red-shifted layer which has H$_{2}$ number densities between 10$^{4}$ to 10$^{6}$ cm$^{-3}$. An estimation for the excitation temperature with an HCO$^+$ column of 10$^{12}$--10$^{13}$ cm$^{-2}$ yield 7--9 K, based on RADEX models. This is sub-thermally excited gas assuming the emitting region is between 15 to 20 K. On the other hand, the rear layer is always thermalized at between 15 to 20 K depending on the physical structure. These values were determined from the typical temperatures at $>$ 500 au in the system, considering the best-fit SED models. Here, we fixed the front layer to 9 K and a rear layer to 15 K. The free parameters are infall velocities $\upsilon_{\rm inf}$, optical depth $\tau_{0}$, and the filling factor $\Phi$. The models also have a broadening factor for the Gaussian line profile. The exact values may change depending on the adopted excitation temperatures. However, the goal of this modelling is to characterize the infalling velocities, which showed a weaker dependence on the adopted excitation temperatures. The systemic cloud velocity in Serpens is taken to be $\sim$8 km s$^{-1}$ (e.g., Kirk et al. 2013). The model profiles are further discussed in Sect.~\ref{bonafide} and Sect.~\ref{classification}.

\subsection{Bonafide Detections}
\label{bonafide}

\noindent {\bf SSTc2d J182902.12+003120.7 (J182902.12):} This is a Class Flat ($\alpha_{IR}$ = 0.27) source with an $L_{\rm bol}$ = 0.05 $L_{\sun}$ (Table~\ref{properties}). The SCUBA-2 maps show a clearly extended object in the 450 $\mu$m band, but less extended emission at 850 $\mu$m (Fig.~\ref{imgs-bf}). There is also a clear detection in the PACS bands. It has been previously detected in the dust continuum emission at 1.1 mm (Enoch et al. 2009), based on which it was categorized as an envelope source (Evans et al. 2009). 

The SED for J182902.12 shows deep absorption in the mid-infrared silicate feature, followed by a steep rise in the far-infrared (Fig.~\ref{seds-bf}). An envelope component is required in order to fit the far-infrared and sub-millimeter points. The best model fit indicates an envelope mass of $\sim$8 $M_{Jup}$ and a disk mass of $\sim$2 $M_{Jup}$, with an edge-on inclination of $\sim$80$\degr$ for the system (Table~\ref{SEDpars}). The HCO$^{+}$ spectrum shows weak emission with a peak at the cloud velocity (Fig.~\ref{mol-bf}).

\vspace{0.05in}

\noindent {\bf SSTc2d J182855.78+002944.8 (J182855.78):} Among the bonafide detections, J182855.78 is the most deeply embedded Class 0 source ($\alpha_{IR}$ = 1.89), with a $L_{\rm bol}$=0.18$L_{\sun}$ (Table~\ref{properties}). This is a bright extended object in the SCUBA-2 maps (Fig.~\ref{imgs-bf}). The extended shape is more clearly seen in the 450 $\mu$m map. There is a point source detection in the PACS 70 $\mu$m and 100 $\mu$m bands, but a marginal detection in the 160 $\mu$m. 

The best model fit to the SED shows a massive envelope of $\sim$0.2 $M_{\sun}$ and a disk component of $\sim$7 $M_{Jup}$, at a $\sim$30$\degr$ inclination for the system (Fig.~\ref{seds-bf}; Table~\ref{SEDpars}). The HCO$^{+}$ (3-2) emission shows a broad profile, with a self-absorption component at the cloud systemic velocity, and a hint of a red-dominated asymmetry (Fig.~\ref{mol-bf}).

\vspace{0.05in}

\noindent {\bf SSTc2d J182949.57+011706.0 (J182949.57):} This is a Class I ($\alpha_{IR}$ = 0.66) system, with an observed $L_{\rm bol}$=0.2 $L_{\sun}$. There is a clear detection in SCUBA-2 and PACS bands (Fig.~\ref{imgs-bf}), and is the first sub-millimeter detection for this YSO. We note that the 160 $\mu$m photometry is affected by the surrounding nebulosity and should be considered as an upper limit. The CanariCam mid-infrared spectrum shows a deep silicate absorption feature (Fig.~\ref{MIRspec}). 

The best SED model fit for J182949.57 shows both an envelope and a disk component (Fig.~\ref{seds-bf}). As seen from the fit, the disk component provides a good fit from the near-infrared up to the far-infrared points including the 10 $\mu$m silicate absorption feature, whereas fitting the sub-millimeter points requires an envelope. We also tested with a model that fits the far-infrared points, but this fit misses the sub-millimeter and mid-infrared photometry. From the best-fit, the envelope and disk mass is estimated to be $\sim$33 $M_{Jup}$ and $\sim$17 $M_{Jup}$, respectively, with a $\sim$40$\degr$ inclination of the system (Table~\ref{SEDpars}). There is no HCO$^{+}$ line observation for this source.

\subsubsection{Mis-identified Sources}
\label{misclassified}

\noindent {\bf SSTc2d J182841.87-000321.3 (J182841.87):} The SCUBA-2 observations provide the first sub-millimeter detection for this Class I object ($\alpha_{IR}$ = 0.37; $L_{\rm bol}$=0.16$L_{\sun}$). There is clear detection for this source at the target location in the PACS 70 $\mu$m and 160 $\mu$m maps (Fig.~\ref{imgs-mis}). This object is at the edge of the PACS 100 $\mu$m map, due to which there is no photometric measurement in this band. 

The SED modeling results for J182841.87 indicate two possible fits (Fig.~\ref{seds-mis}). The best-fit model provides a good fit to the sub-millimeter points as well as the full depth of the 10$\mu$m silicate feature, but misses the mid- and far-infrared points. The second model fits the mid- and far-infrared photometry, but misses the sub-millimeter points as well as the silicate feature. This indicates that a small envelope component is required in order to fit the sub-millimeter points. The best-fit is obtained for an envelope mass of $\sim$13 $M_{Jup}$, a disk mass of $\sim$6 $M_{Jup}$, and an inclination of 40$\degr$ for the system. In contrast, the second fit is mainly dominated by the disk emission, and is obtained for a close to edge-on disk at an inclination of $\sim$69$\degr$ and a mass of $\sim$7 $M_{Jup}$. There is no HCO$^{+}$ (3-2) line detection. The characteristics for J182841.87 are consistent with either a Stage II edge-on disk source, or a Stage I/II transition object with a tenuous envelope. This object has been mis-identified as an embedded Class I system.

\vspace{0.05in}

\noindent {\bf SSTc2d J182956.67+011239.2 (J182956.67):} This is a Class Flat ($\alpha_{IR}$ = -0.11) source with observed $L_{\rm bol}$=0.06 $L_{\sun}$. There is a point source detection in the SCUBA-2 maps, but no clear point source is seen in the PACS maps (Fig.~\ref{imgs-mis}). The PACS photometry should thus be considered as the upper limits obtained at the target position. 


All possible SED model fits for this source indicate a disk-dominated object at an edge-on inclination of $\sim$80$\degr$, with a disk mass of $\sim$12 $M_{Jup}$ (Fig.~\ref{seds-mis}; Table~\ref{SEDpars}). There is an indication of an extremely tenuous envelope component of $\sim$0.4 $M_{Jup}$ mass. The HCO$^{+}$ (3-2) line shows a broad profile centered close to the cloud velocity, with a hint of self-absorption at higher velocities (Fig.~\ref{mol-bf}). The Stage 0/I classification for J182956.67 by Heiderman \& Evans (2015), based on HCO$^{+}$ detection, is inconsistent with the SED modeling results. It may be the case that there are different kinds of sources in the beam, such as, a faint star superposed on a starless core, resulting in a strong detection in the HCO$^{+}$ line. An $\alpha_{IR}$ of -0.11 places it closer to the Class Flat/Class II boundary (-0.3). It may possess a very tenuous envelope, if any, and is more likely to be a case of a Stage II or Class II edge-on disk source.

\vspace{0.05in}

\noindent {\bf SSTc2d J182902.84+003009.6 (J182902.84)}: This is a Class Flat ($\alpha_{IR}$=-0.14) source with observed $L_{\rm bol}$=0.17 $L_{\sun}$. There is a point-like detection for this object in the SCUBA-2 and PACS 70 $\mu$m and 100 $\mu$m bands, though it appears (slightly) extended in the SCUBA-2 450\,$\mu$m map (Fig.~\ref{imgs-mis}). There is no detection in the PACS 160 $\mu$m band. A detection at 1.1 mm has been previously reported, based on which it was categorized as an envelope source (Enoch et al. 2009; Evans et al. 2009). 

\begin{figure}
  \centering              
     \includegraphics[width=2.5in]{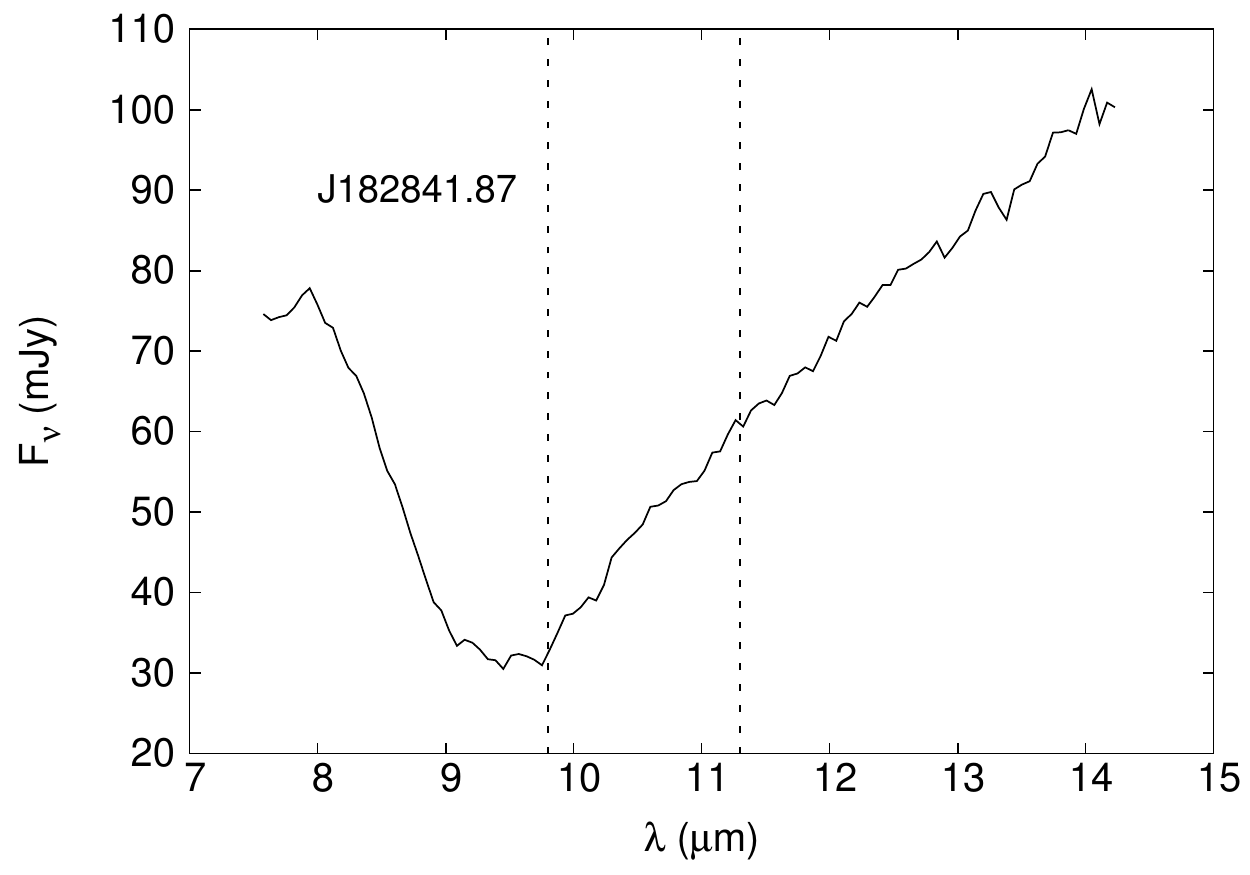} 
     \includegraphics[width=2.5in]{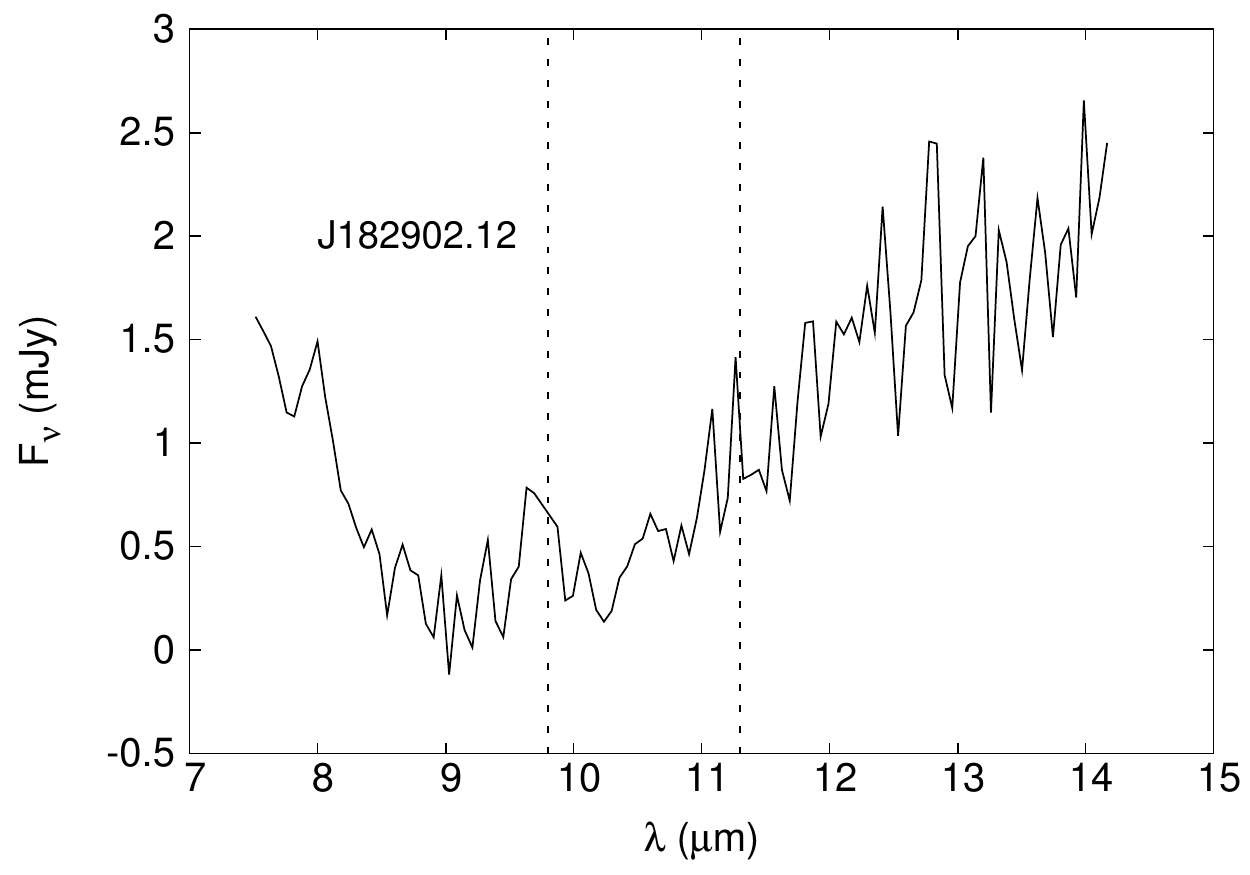} 
     \includegraphics[width=2.5in]{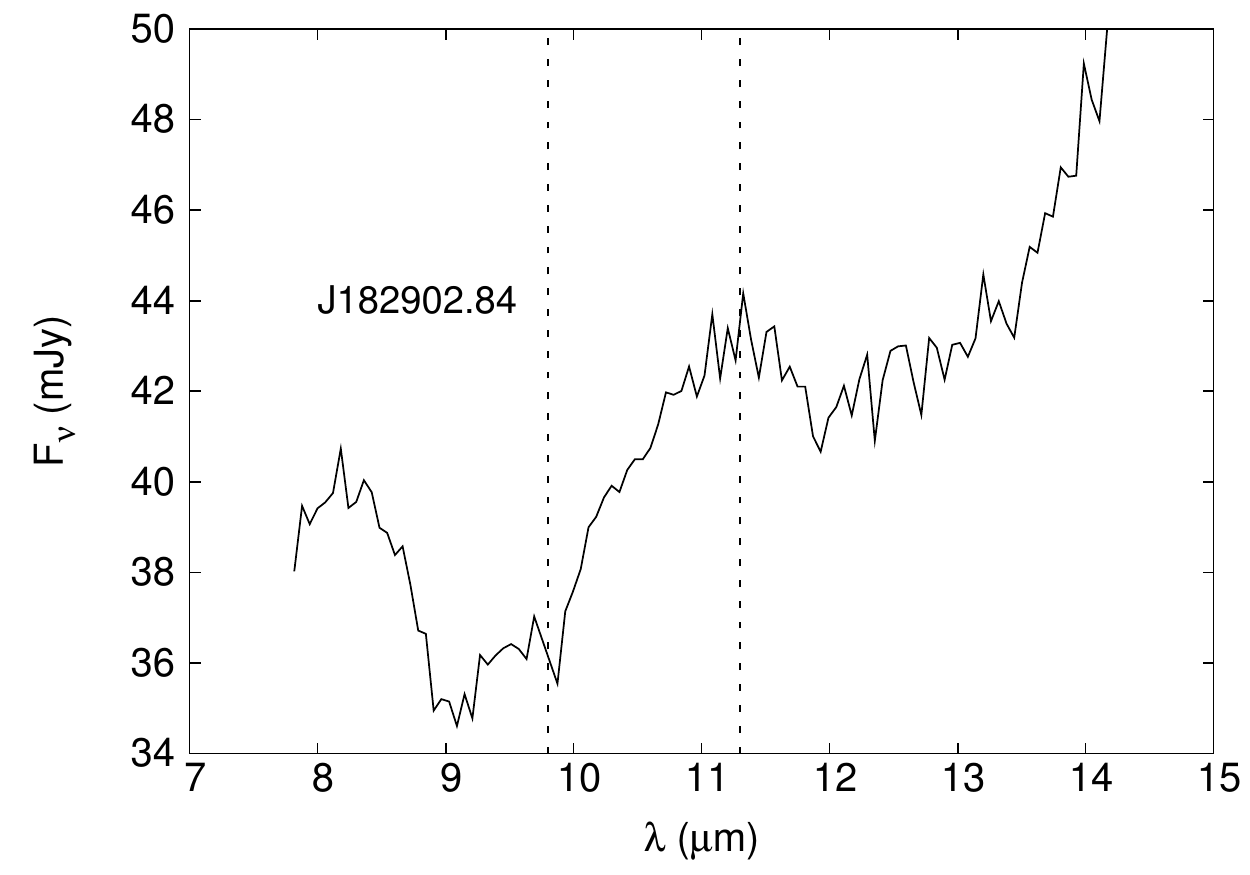} 
     \includegraphics[width=2.5in]{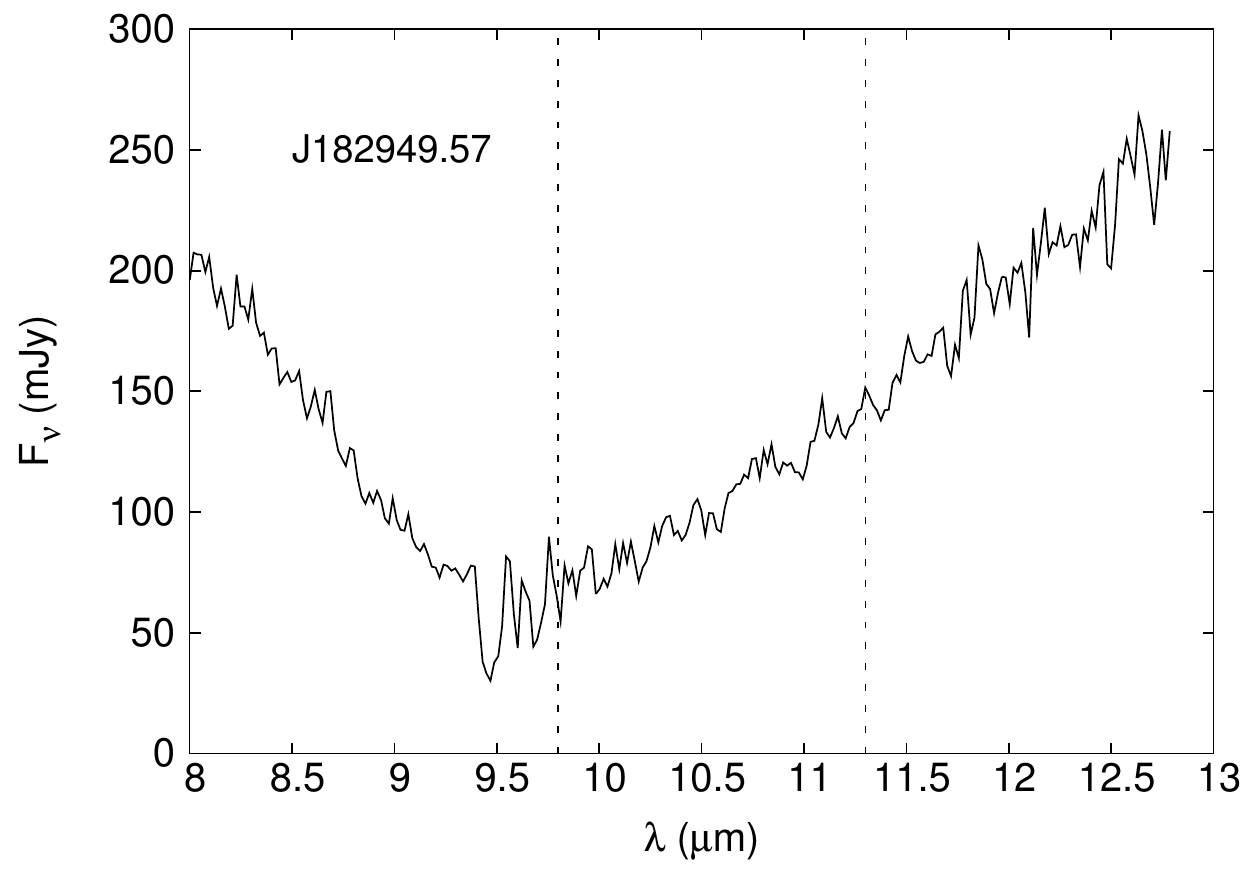} 
     \includegraphics[width=2.5in]{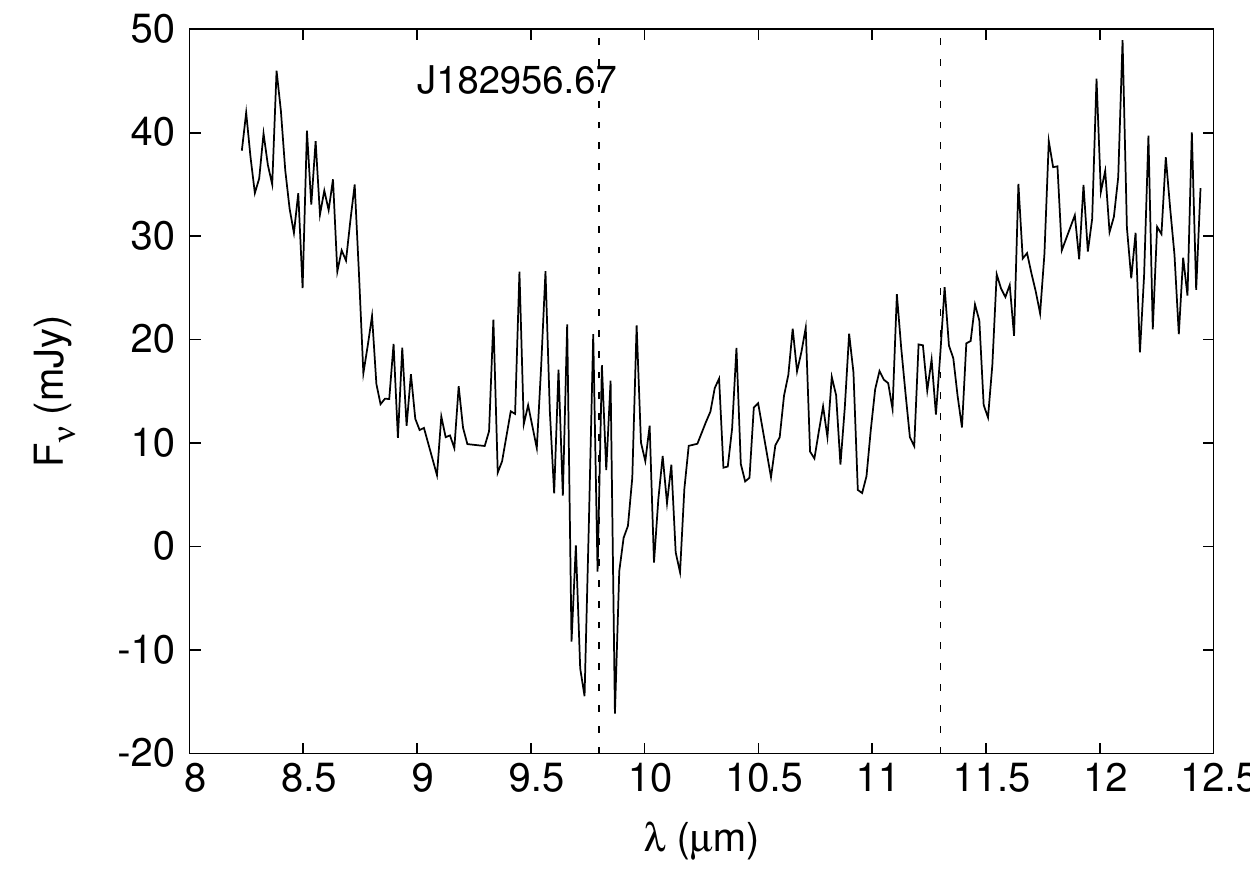} 
     \includegraphics[width=2.5in]{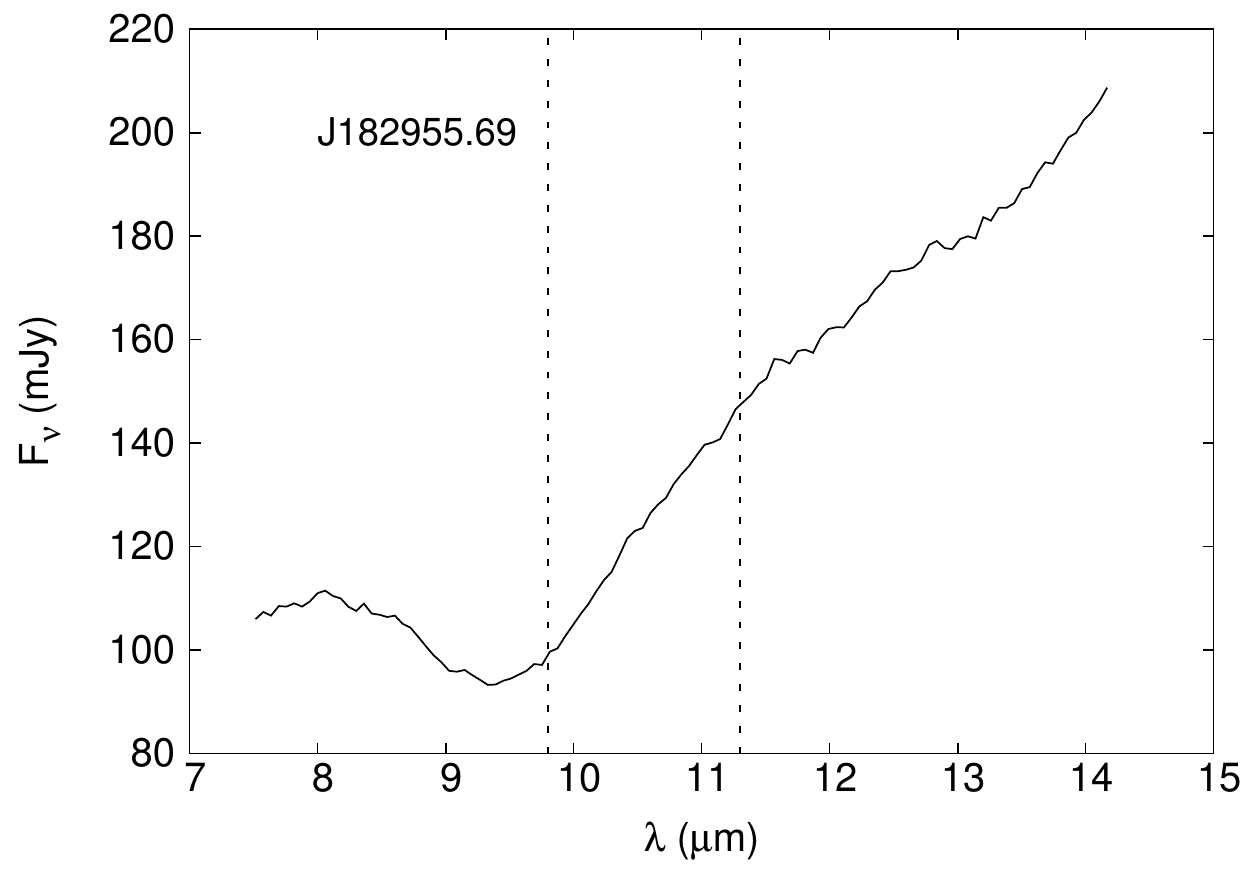} 
     \caption{The mid-infrared spectra. The vertical lines mark the amorphous olivine and crystalline forsterite peaks at 9.8 $\mu$m and 11.3 $\mu$m, respectively. }     
     \label{MIRspec}
  \end{figure}

J182902.84 shows an interesting 10$\mu$m silicate feature, with both an emission and an absorption component (Fig.~\ref{MIRspec}). Such a feature has been earlier noted in a brown dwarf edge-on disk (e.g., Luhman et al. 2007). The silicate emission component has an origin in the optically thin surface layers of the disk, whereas the absorption is caused by the highly extinguished inner disk wall. The best model fit to the J182902.84 SED shows an edge-on disk at a $\sim$75$\degr$ inclination, with a mass of $\sim$11 $M_{Jup}$ (Fig.~\ref{seds-mis}). No envelope component is required, and a disk alone is adequate to fit the sub-millimeter points. There are no HCO$^{+}$ (3-2) observations available for this source. Overall, J182902.84 appears to be a case of a mis-identified object, showing characteristics that are more similar to a disk-only Stage II or Class II source. 

\vspace{0.05in}

\noindent {\bf SSTc2d J182955.69+011431.6 (J182955.69)}: This is a Class Flat source ($\alpha_{IR}$ = -0.23), with an observed $L_{\rm bol}$=0.2 $L_{\sun}$. The spectral type determined from near-infrared spectroscopy is M4.5$\pm$0.5 (Gorlova et al. 2010). J182955.69 shows a bright point-like detection in the SCUBA-2 maps, the first sub-millimeter detection for this object (Fig.~\ref{imgs-mis}). It is difficult to confirm a detection in the PACS map as it lies in a nebulous, confused region. Thus the PACS photometry should be considered as upper limits. Results from SED modeling for J182955.69 indicate an edge-on disk at $\sim$75$\degr$ inclination, with a disk mass of $\sim$18 $M_{Jup}$ (Fig.~\ref{seds-mis}). The best-fit also indicates a very weak envelope component of $\sim$0.1 $M_{Jup}$. The spectral slope lies close to the Class Flat/Class II boundary. J182955.69 is thus a mis-identified object, and appears as a disk-dominated Stage II or Class II source.

\subsection{Possible proto-brown dwarf candidates}
\label{pbds}

For 6 sources in our sample, there is no detection (SNR$<$2-$\sigma$) in the SCUBA-2 maps. None of these sources are detected in the PACS maps. The observed $L_{\rm bol}$ for these objects is $<$0.04 $L_{\sun}$, implying that the envelope masses are too low, and the non-detection can be explained by the poor sensitivity both in the SCUBA-2 and PACS maps. We consider these objects as potential proto-brown dwarf candidates. 

Among these candidates, four sources show emission in the HCO$^{+}$ (3-2) line (Fig.~\ref{mol-nd}). The line emission for three sources show a broad profile with self-absorption at the systemic velocity. For the source J182852.76+002846.8, the HCO$^{+}$ emission shows slightly blue-shifted asymmetry indicative of an infalling envelope. A weak indication of a blue component is also seen for J182959.03+011225.1, for which the main emission with a peak at $\sim$8 km s$^{-1}$ must be dominated by the foreground cloud material. The line appears to be a blend of multiple velocity components, possibly arising from a molecular outflow. In contrast, the source J182947.01+011626.9 shows a slightly red-dominated asymmetry, as observed in some disk-dominated objects (e.g., van Kempen et al. 2009), although such asymmetries could also appear in some embedded sources driving an outflow (e.g., Harsono et al. {\it in prep}). All four of these sources have been categorized as Stage 0/I protostars due to HCO$^{+}$ line detection (Heiderman \& Evans 2015). We note that while Heiderman \& Evans (2015) have searched for HCO$^{+}$ observations within 14$\arcsec$ of the {\it Spitzer} c2d source position, all four of these sources with HCO$^{+}$ line detection are located at a $\sim$20$\arcsec$ separation from a bright source in the SCUBA-2 maps. Since the nearby bright object is within the CSO 30$\arcsec$ beamsize, it may have contaminated the observed line emission. Spatially resolved observations at higher sensitivities can provide better insight into their characteristics.

\section{Discussion}
\label{discussion}

\subsection{Classification of Very Low-Mass/Sub-stellar YSOs}
\label{classification}

The traditional method of determining the evolutionary class of a young stellar object (YSO) is to measure the near- to mid-infrared spectral index, $\alpha_{IR}$ = {\it d} log($\lambda F_{\lambda}$)/{\it d} log($\lambda$), as first defined by Lada \& Wilking (1984) and Adams et al. (1987). The typical wavelength range considered to measure $\alpha_{IR}$ is 2--24$\mu$m. The now widely used classification scheme consists of 5 classes of YSOs: Class 0 is the earliest, deeply embedded stage with a high sub-millimeter to bolometric luminosity ratio (And\'{r}e et al. 1993); Class I are more evolved embedded YSOs with $\alpha_{IR}$ $>$ 0.3; the `Flat Spectrum' sources (-0.3 $<$ $\alpha_{IR}$ $<$ 0.3; Greene et al. 1994) are at an intermediate stage between Class I and II and have tenuous envelopes compared to Class I objects; Class II (-2 $<$ $\alpha_{IR}$ $<$ -0.3) are T Tauri stars with gas-rich circumstellar disks but no envelope material; and Class III ($\alpha_{IR}$ $<$ -2) are pre-main sequence stars which may possess tenuous disk material. Another classification method is based on the bolometric temperature, $T_{bol}$, which was first connected to the classes defined by $\alpha_{IR}$ by Chen et al. (1995). In Table~\ref{properties} (column 8), we have listed the classification as originally determined by Evans et al. (2009), using both the spectral index and $T_{bol}$ criteria. The classes from the two schemes are similar, except for the transition cases close to the boundary between Class Flat and Class II sources. 

From our analysis based on SED modeling presented in Sect.~\ref{misclassified}, we have found 4 out of 7 bonafide detections to be mis-identified objects, such that their characteristics do not comply with the original classification based on $\alpha_{IR}$ or $T_{bol}$ criteria. These include 3 Class Flat sources that show characteristics similar to Class II objects, and a Class I system that appears to be a Class Flat source. The revised classification based on SED modeling results are listed in Table~\ref{properties} (column 9). As noted by Whitney et al. (2003), there can be a wide range in $\alpha_{IR}$ for a given $T_{bol}$ category, and therefore classifying a YSO only based on its observed characteristics of the spectral slope and/or the bolometric temperature can be erroneous, particularly for cases such as a face-on embedded YSO which may be mis-identified as Class II, or an edge-on flaring disk which may be mis-identified as a Class I object. 

The ``Stage'' classification scheme introduced by Whitney et al. (2003) and later modified by Robitaille et al. (2006), is based on the evolutionary stage of a YSO using its true physical characteristics, such as the disk mass and envelope accretion rate, regardless of the observed properties. In this classification scheme, Stage 0/I objects have significant infalling envelopes, with $\dot{M}_{env}$/$M_{star}$ $>$ 10$^{-6}$ yr$^{-1}$, Stage II sources have tenuous envelopes but gas-rich optically thick disks, with $\dot{M}_{env}$/$M_{star}$ $<$ 10$^{-6}$ yr$^{-1}$ and $M_{disk}$/$M_{star}$ $>$ 10$^{-6}$, while Stage III objects with $\dot{M}_{env}$/$M_{star}$ $<$ 10$^{-6}$ yr$^{-1}$ and $M_{disk}$/$M_{star}$ $<$ 10$^{-6}$ have no envelopes but may have tenuous disks. Some additional constraints are: Stage 0 objects have $M_{disk}$/$M_{env}$ $<<$ 1 and $M_{circum}$/$M_{star}$ $\sim$ 1, Stage I have 0.1 $<$ $M_{disk}$/$M_{env}$ $<$ 2 and $M_{circum}$ $<$ $M_{star}$, while Stage II have $M_{env}$=0 and $M_{disk}$/$M_{star}$ $<<$ 1. Here, $M_{circum}$ refers to the envelope+disk mass. An intermediate phase is the Stage I Transition (Stage I-T) category (e.g., van Kempen et al. 2009), with $\dot{M}_{env}$/$M_{star}$ $\sim$(1-2)$\times$10$^{-6}$ yr$^{-1}$ and $M_{disk}$/$M_{env}$ $>$ 2. These sources have tenuous envelopes compared to the early Stage I objects.

More recently, a classification scheme combining dust continuum and molecular line observations, notably the high gas density tracer HCO$^{+}$ molecule, has been developed to identify the truly embedded Class 0/I sources and separate them from Class II disk sources (e.g., van Kempen et al. 2009). In addition to a sub-millimeter dust continuum detection peaking on-source, this criterium is based on the strength of the HCO$^{+}$ emission; in particular, the integrated intensity of the HCO$^{+}$ (4-3) line being $>$0.4 K km s$^{-1}$ is considered to be a good metric for a Class 0/I or Stage 0/I sources, while the HCO$^{+}$ line is either undetected or detected at a very weak level ($T_{mb}$ $<$ 0.1 K) in disk-dominated Class II/Stage II sources. Another notable feature is an axisymmetrical profile observed in the HCO$^{+}$ line, which may be indicative of an infalling envelope. Typically, a blue-dominated infall asymmetry has been seen in the HCO$^{+}$ profile of some low-mass protostars, while a red-dominated axisymmetry is usually associated with disk-only objects or the presence of an outflow (e.g., Thi et al. 2004; Evans et al. 1999; van Kempen et al. 2009). However, this is not always the case, since red-dominated asymmetry has been observed in embedded sources (Harsono et al. {\it in prep.}), an interpretation of which could be an expanding envelope due to an outflow.

\begin{figure}
     \includegraphics[width=2.7in]{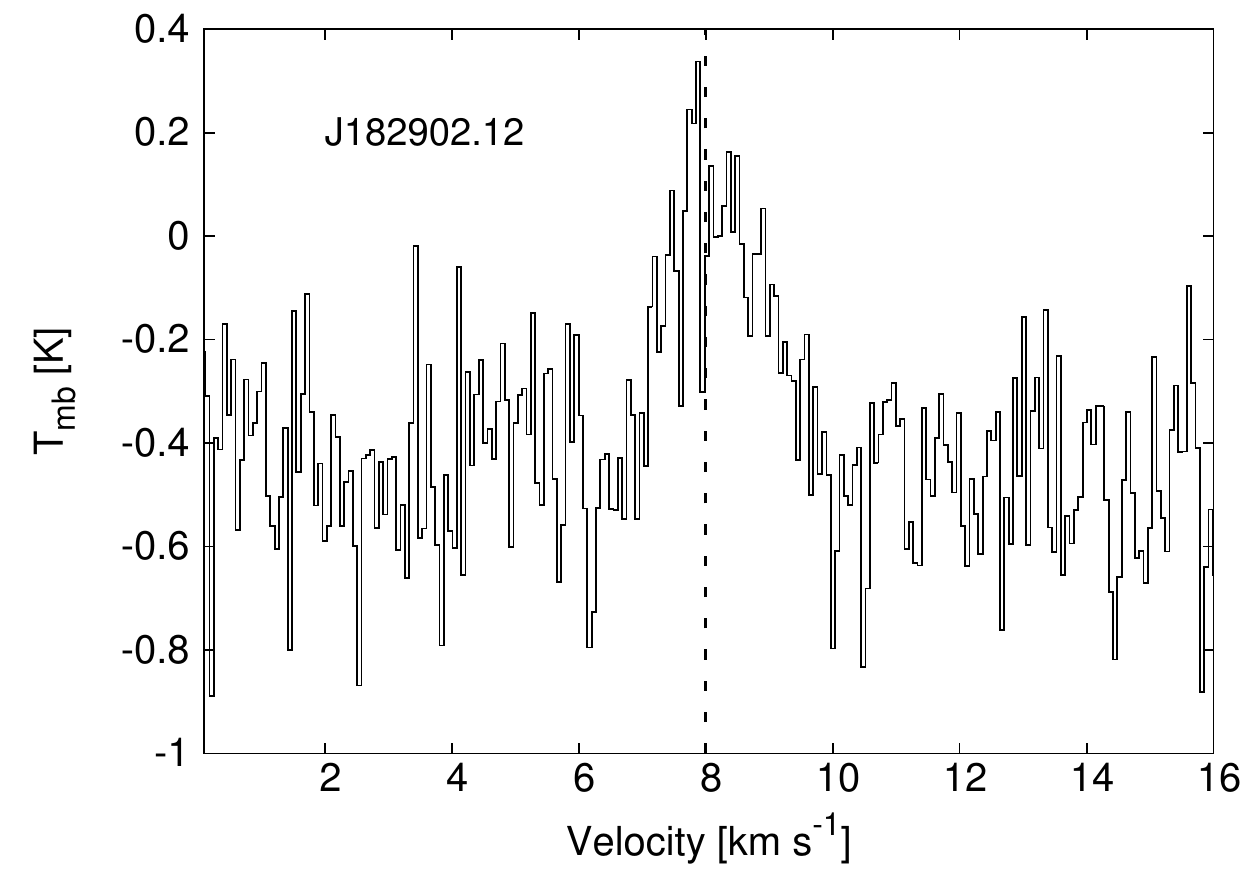}     
     \includegraphics[width=2.3in]{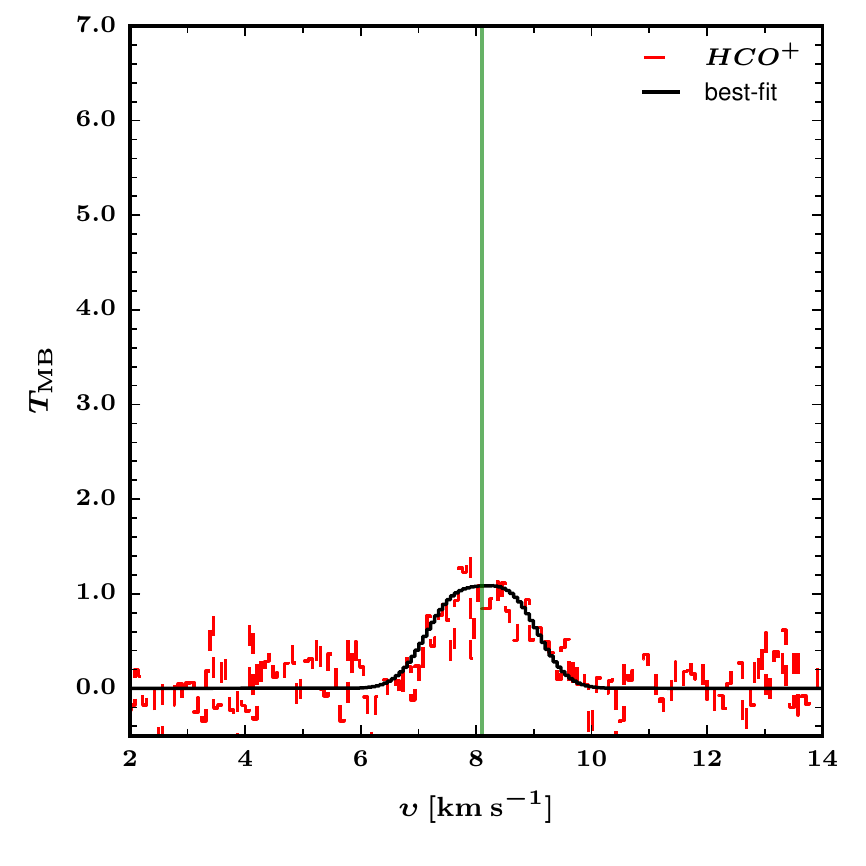}     \\
     \includegraphics[width=2.7in]{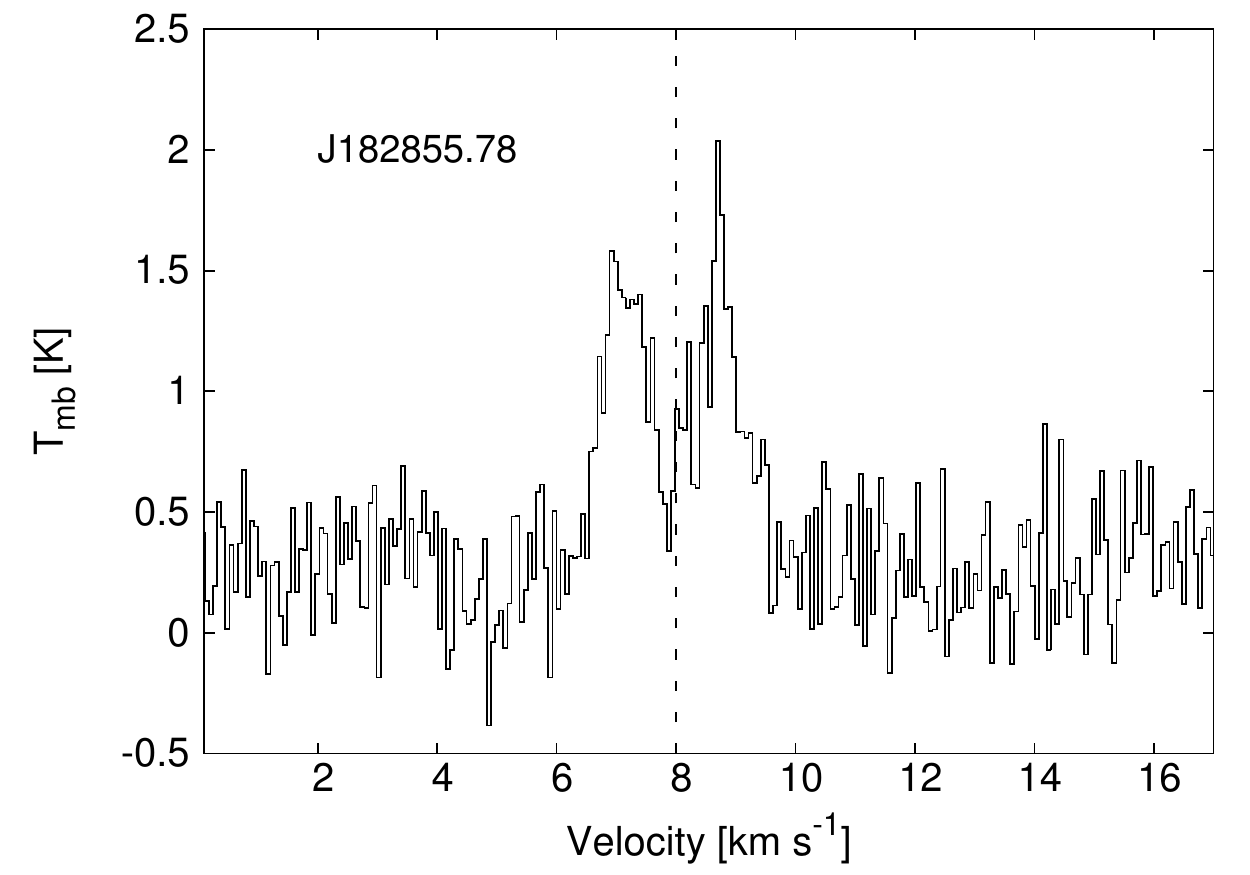}    
     \includegraphics[width=2.3in]{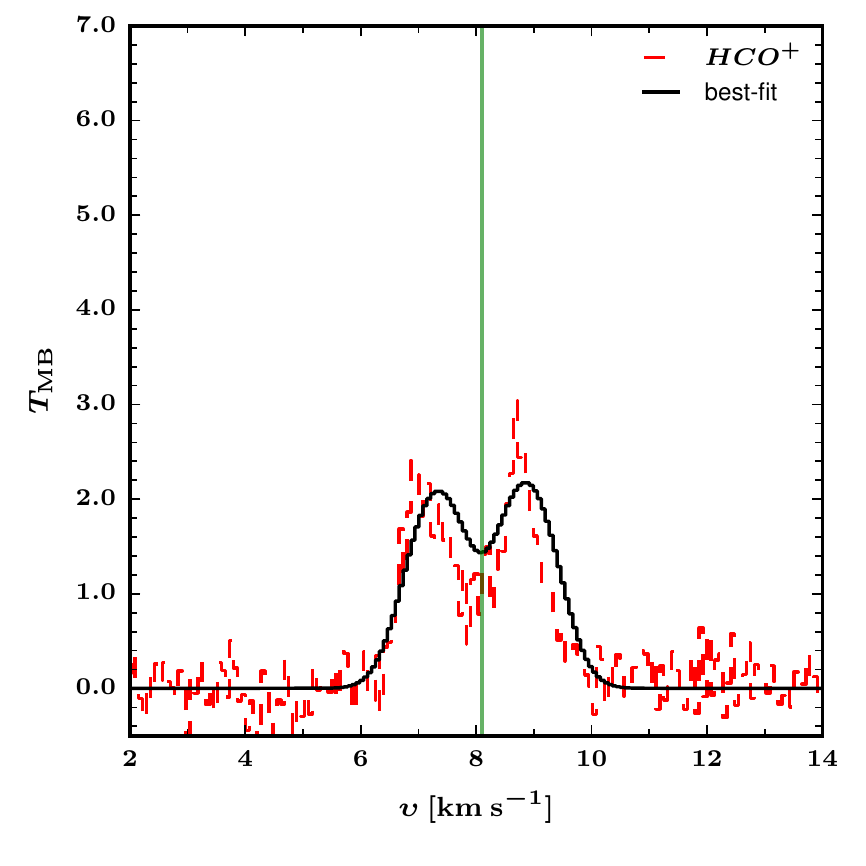}       \\        
     \includegraphics[width=2.7in]{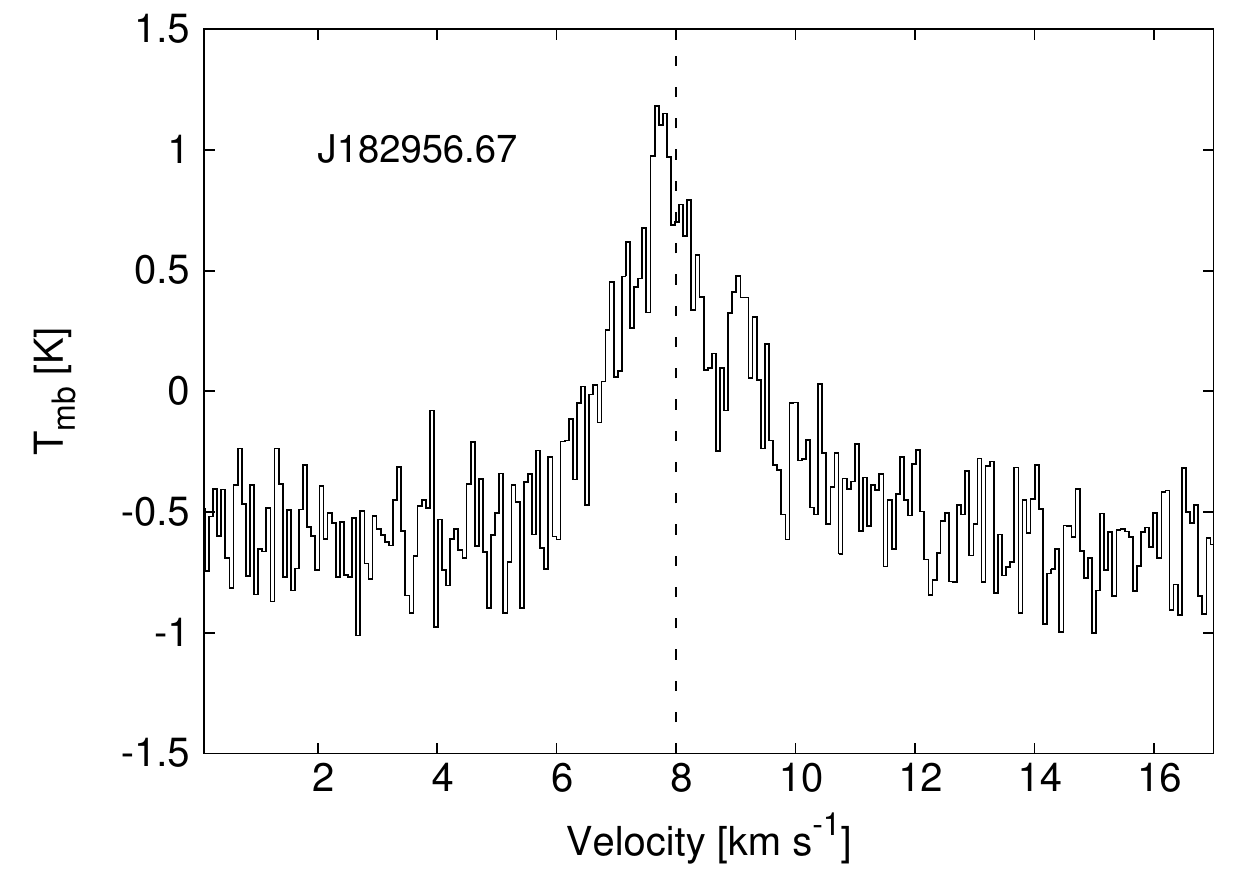}        
     \includegraphics[width=2.3in]{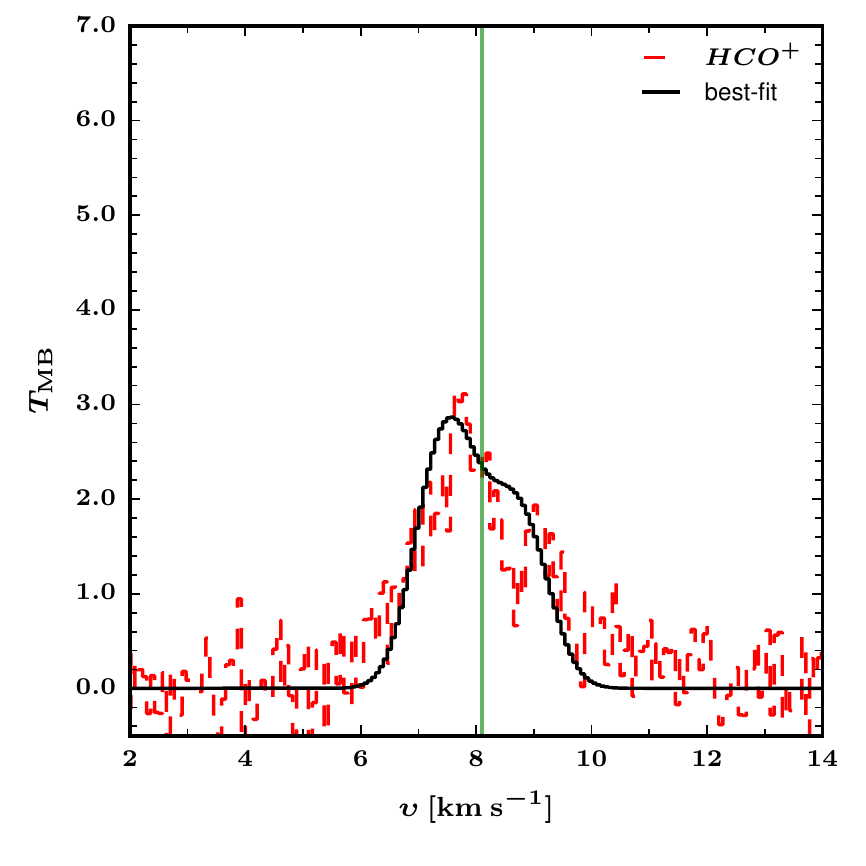}          
     \caption{HCO$^{+}$ (3-2) observed (left) and model (right) line profiles for the bonafide detections. The vertical line marks the systemic cloud velocity.  }
     \label{mol-bf}
  \end{figure}

Among the bonafide detections in our present work, J182902.12, J182855.78, and J182956.67 have previously been categorized as Stage 0/I systems by Heiderman \& Evans (2015), based on the criteria of HCO$^{+}$ (3-2) integrated line intensity of $\geq$0.68 K km s$^{-1}$, while J182841.87 has been categorized as a Stage II system. We note that the Heiderman \& Evans (2015) criteria is solely based on a HCO$^{+}$ detection within 14$\arcsec$ of the {\it Spitzer} c2d source position. As can be seen in Fig.~\ref{mol-bf}, J182902.12 show a broad HCO$^{+}$ profile centered at the cloud velocity. The line model fit shows a typical Gaussian, which implies that there is no infalling envelope in this case. We have shown through SED modeling that J182902.12 is a Class Flat system and may possess a tenuous envelope. Using the ``Stage'' classification scheme also places J182902.12 in the Stage I-T category. Therefore, the Stage 0/I classification based on the HCO$^{+}$ detection is inconsistent with other schemes. For the case of J182855.78, results from line modeling indicate that the observed HCO$^{+}$ asymmetry can be explained by an infalling envelope model (Fig.~\ref{mol-bf}). The SED modeling results and the Stage classification also indicate that this is a genuine case of a Class 0/I or Stage 0/I system. For the case of J182956.67, the HCO$^{+}$ profile shows a hint of a blue-shifted asymmetry and self-absorption at higher velocities (Fig.~\ref{mol-bf}). Results from line modeling for J182956.67 suggest that there may be blue-dominated infall asymmetry due to a weak envelope component with velocities between 0.5 -- 1 km s$^{-1}$ and a mass infall rate of $<$ 10$^{-7}$ $M_{\sun}$ yr$^{-1}$. However, both the SED model fit and the Stage classification indicate that this is a Class II edge-on disk system or a Stage II source, rather than embedded Stage 0/I systems (Table~\ref{properties}; column 9). On the contrary, J182841.87 can be placed in the Stage I-T category, and shows signatures of a tenuous envelope (Sect.~\ref{model}), unlike the Stage II classification based on HCO$^{+}$ non-detection. The envelope mass in this system is too low to produce any detectable emission in a high-density tracer such as the HCO$^{+}$ (3-2) line. The sub-millimeter emission observed towards this source likely arises from a tenuous dusty envelope with low HCO$^{+}$ abundance. Overall, there appears to be less consistency in the classification determined from the physical structure of the system, and that from the molecular line detection (Table~\ref{properties}; cols. 8, 9).

\begin{figure}
  \centering           
     \includegraphics[width=3in]{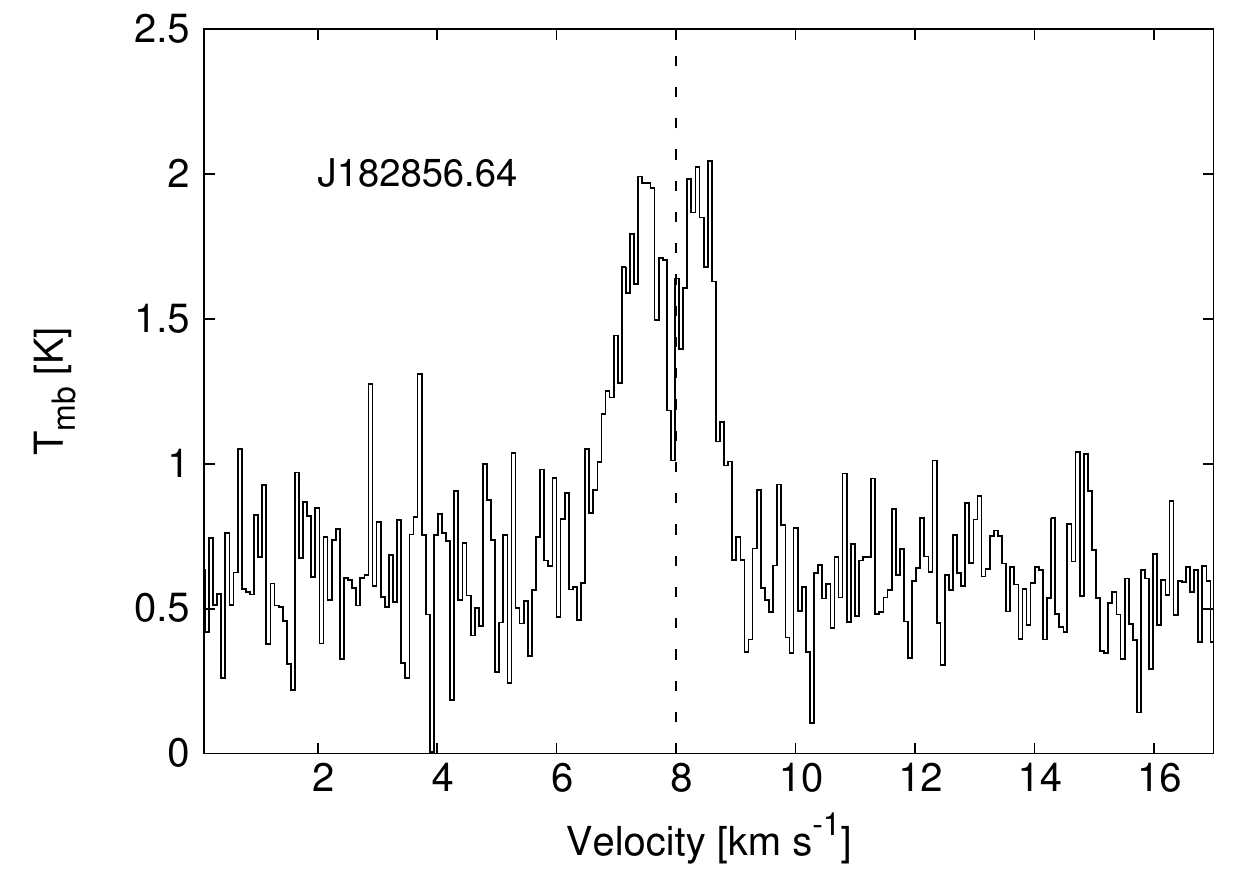}     
     \includegraphics[width=3in]{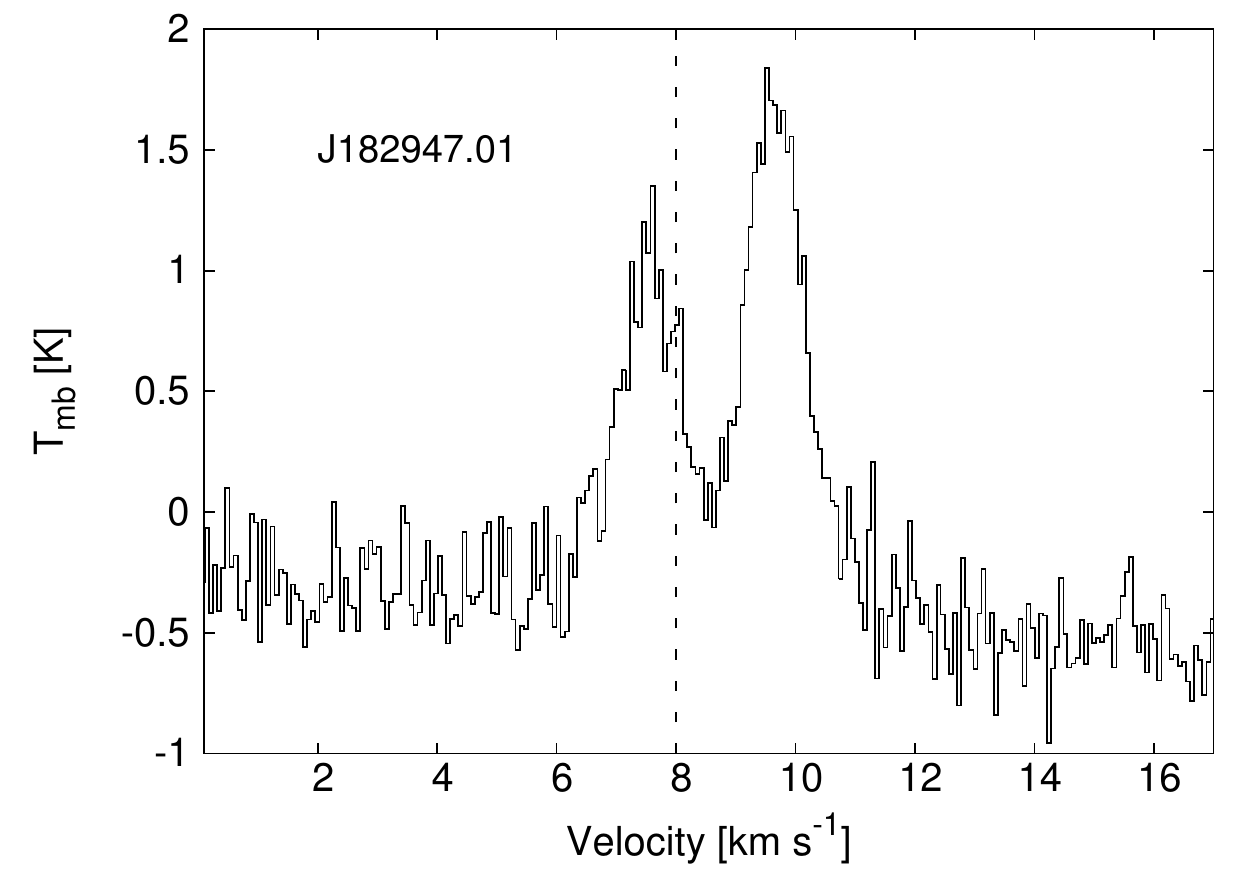}       
     \includegraphics[width=3in]{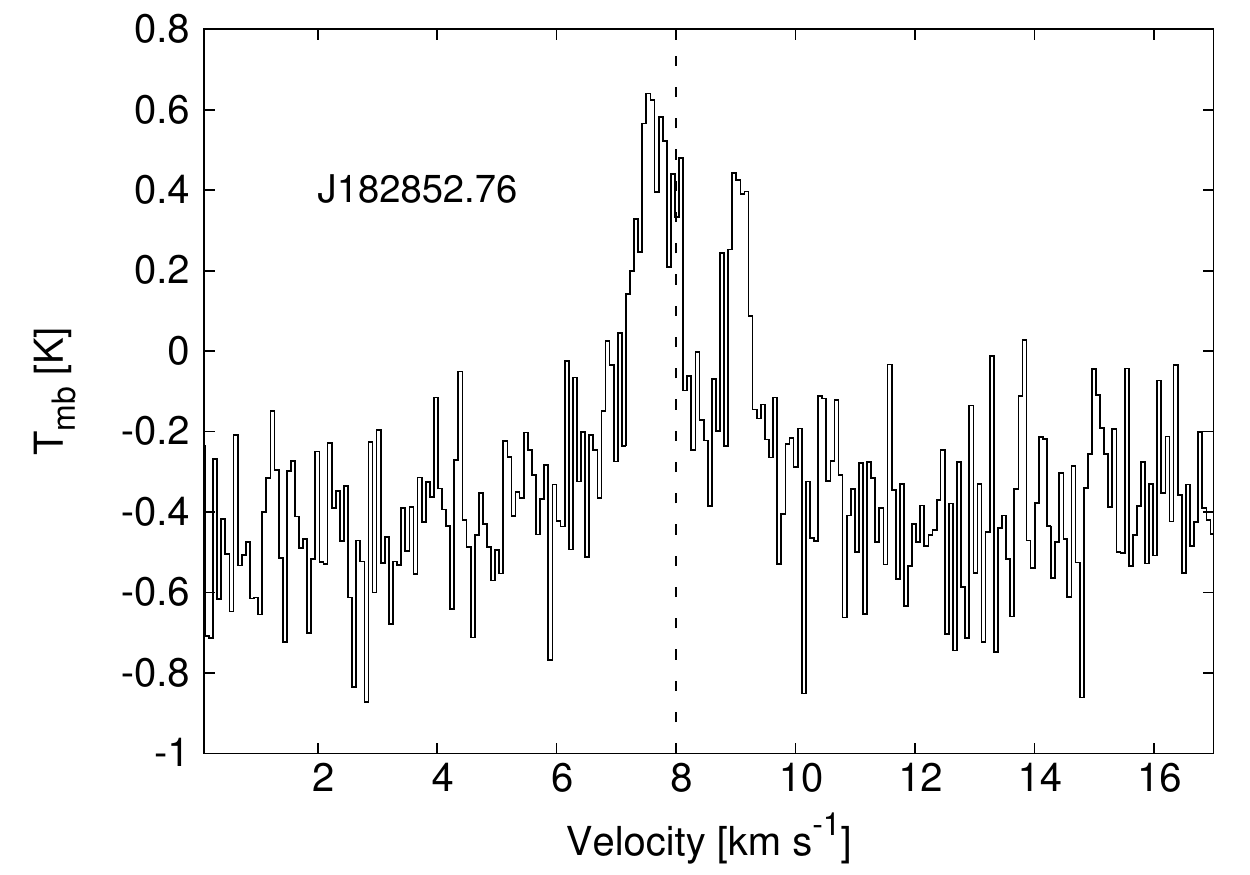}   
     \includegraphics[width=3in]{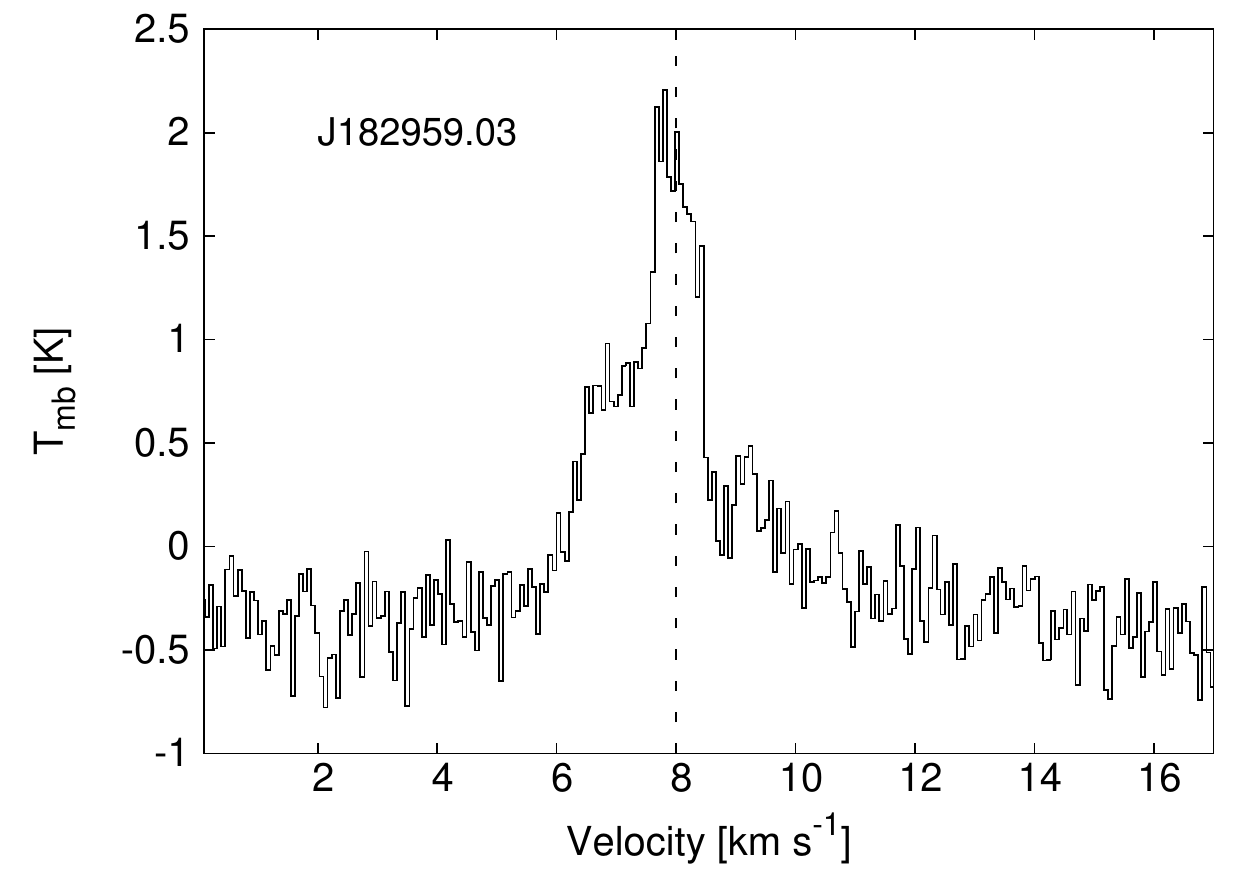}        
     \caption{HCO$^{+}$ (3-2) line profiles for the proto-brown dwarf candidates that are undetected in the SCUBA-2 maps (Sect.~\ref{pbds}). The vertical line marks the systemic cloud velocity.  }
     \label{mol-nd}
  \end{figure}

It is important to note that there are shortcomings with these various classification schemes; none of these schemes have been developed using very low-luminosity objects. Such low-luminosity Stage 0/I VLM/BDs are expected to have smaller disks and envelopes than typical protostars (e.g., White \& Hillenbrand 2004), which could result in a Stage I-T or a Stage II classification. Since these classification schemes do not extend to the sub-stellar mass regime, there may be a potential bias in classifying bonafide Stage 0/I or Class 0/I brown dwarfs as Stage II or Stage I-T objects. A revision of the criteria for the HCO$^{+}$ integrated line intensity of $>$0.4 K km s$^{-1}$ is also required for objects at the very low-mass end. For cases such as the proto-brown dwarf candidate J182852.76+002846.8 (Sect.~\ref{pbds}), the HCO$^{+}$ emission is quite weak, but shows a slightly blue-shifted asymmetry that suggests the presence of an infalling envelope (Fig.~\ref{mol-nd}). As mentioned, using these various schemes to constrain the evolutionary stage of the system, we have found 4 out of 7 bonafide detections to be mis-identified objects. Assuming that the 6 proto-brown dwarf candidates undetected in sub-millimeter continuum (Sect.~\ref{pbds}), and the marginal/unresolved sources (Sect.~\ref{appendix}) are all correctly classified, we have a $\sim$20\% fraction of mis-identified objects among Class 0/I/Flat VLM/BD sources in Serpens. In comparison, Heiderman \& Evans (2015) estimate a $\sim$16\% fraction of the full Class 0/I/Flat sample from Evans et al. (2009) in Serpens to be mis-identified, and suggest these to be Stage II disk-dominated objects. As we have shown in Sect.~\ref{misclassified}, at least 4 of the cases classified as Stage 0/I objects in Heiderman \& Evans (2015) are more evolved cases of Stage II/I-T objects. Thus in both of these studies, the estimate on the mis-identified fraction is likely to be a lower limit.

\subsection{Total Mass of the System}
\label{totalmass}

For the bonafide detections in the SCUBA-2 maps, we have derived the total (dust+gas) mass arising from the (envelope+disk) components of the system, $M^{850}_{\rm{g+d}}$, using the 850\,$\mu$m flux density. These masses have been derived assuming a dust temperature, $T_{dust}$, of 10 K, a gas-to-dust mass ratio of 100, and a dust mass opacity coefficient at 850 $\mu$m of 0.0175 cm$^{2}$ gm$^{-1}$ (column 6, Table 1 in Ossenkopf \& Henning 1994, corresponding to agglomerated dust grains with thin ice mantles at densities $\sim$10$^{6}$ cm$^{-3}$). Table~\ref{masses} lists the $M^{850}_{\rm{g+d}}$ estimates, along with the (envelope+disk) mass (dust+gas) obtained from the best model fit to the observed SED, $M^{\rm SED}_{\rm{g+d}}$. 

For all cases except J182855.78, $M^{850}_{\rm{g+d}}$ is much higher than $M^{\rm SED}_{\rm{g+d}}$. This suggests that either $T_{dust}$ is higher than the assumed value of 10 K, or there may be more flux in the 850 $\mu$m beamwidth than arising from the compact source itself. We have estimated the value for $T_{dust}$ that provides the best match between $M^{850}_{\rm{g+d}}$ and $M^{\rm SED}_{\rm{g+d}}$ (Table~\ref{masses}; column 4). A higher $T_{dust}$ value of $\sim$20-40 K provides a better match between the two estimates. Considering that we have used the masking technique that should minimize the effects of beam dilution (Sect.~\ref{submm-obs}), the system temperature can be expected to be warmer than 10 K. The high $T_{dust}$ values are consistent with the average dust temperatures of $\sim$30-40 K found from radiative transfer models of low-mass Class I protostars (e.g., Jorgensen et al 2011; Harsono et al. 2015). Also listed in Table~\ref{masses} (columns 5, 6) is the envelope and/or disk size obtained from the SED fit ($S_{model}$), and the spatial extent of the source as measured in the 850 $\mu$m image ($S_{obs}$). As can be seen, $S_{obs}$ is $\sim$2-10 times larger than the size $S_{model}$ from the SED fit. This can be expected given the large beamsize in single-dish observations, and $S_{obs}$ is likely an upper limit on the actual size of the system; interferometry can provide a more robust measurement on the source size. The larger than expected $S_{obs}$ suggests that $M^{850}_{\rm{g+d}}$ could be overestimated, and should be considered as an upper limit to the total (cold) mass in the system.

The derived values of $M^{\rm SED}_{\rm{g+d}}$ represent the available mass reservoir in the disk and envelope in the form of gas and dust. Low values of $M^{\rm SED}_{\rm{g+d}}$, such as for J182902.12 and the last four objects in Table~\ref{masses}, imply that the mass of the central objects in these systems will not increase appreciably in the subsequent evolution. This conjecture is reinforced by the fact that they are classified as Flat objects, meaning that most of $M^{\rm SED}_{\rm{g+d}}$ is contained in the disk rather than in the envelope.

We now proceed with setting constrains on the mass of the central object using the measured bolometric luminosities and numerical simulations of stellar evolution described in Baraffe et al. (2012) and Vorobyov et al. (2016). These authors calculate stellar properties starting from a protostellar seed of $1.0~M_{\rm Jup}$, and using the realistic mass accretion rates derived from numerical hydrodynamics simulations of disk evolution. While Baraffe et al. use precalculated mass accretion rates, Vorobyov et al. employ a fully self-consistent coupled evolution of the central object with disk hydrodynamic models. In both cases, the Lyon stellar evolution code is used to calculate the stellar properties (Chabrier \& Baraffe 2000). We refer the reader to these works for a detailed description. Here, we simply provide the results of the numerical simulations by Vorobyov et al. (2016).

\begin{figure}
  \centering              
     \includegraphics[width=5in]{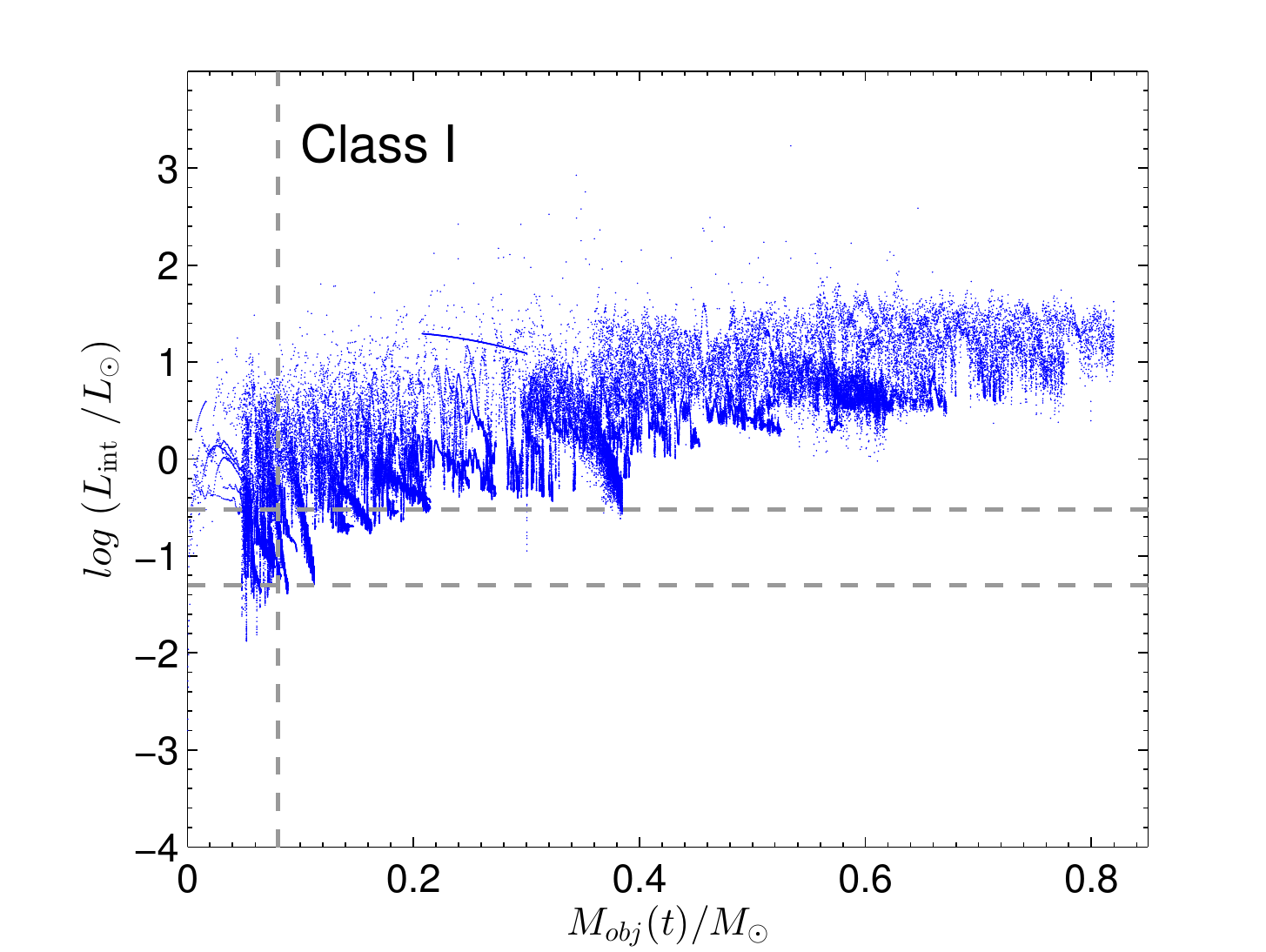} 
     \caption{Internal luminosity ($L_{\rm int}$) vs. central object mass ($M_{\rm obj}$) diagram obtained using 31 models of accreting brown dwarfs and protostars from numerical hydrodynamics simulations of Vorobyov et al. (2016). The data correspond to the Class I phase of stellar evolution. The horizontal dashed lines indicate the minimum and maximum bolometric luminosities of $0.05~L_\odot$ and $0.3~L_\odot$ in our sample. The vertical line separating the sub-stellar and stellar regimes. }
     \label{fig:Lum}
  \end{figure}

Figure~\ref{fig:Lum} presents the $L_{\rm int}$-–$M_{\rm obj}$ diagram for 31 models of accreting proto-brown dwarfs and protostars, where $L_{\rm int}$ is the sum of accretion and photospheric luminosities of the central object (star or brown dwarf) and $M_{\rm obj}$ is the current mass of the central object. The data correspond to the Class I phase of stellar evolution. The horizontal dashed lines indicate the minimum and maximum bolometric luminosities of $0.05~L_\odot$ and $0.3~L_\odot$ in our sample (see Table~\ref{properties}). We note that the bolometric luminosity should be considered as an upper limit on the true internal luminosity, $L_{\rm int}$, of the system. The $L_{int}$ is the luminosity of the central protostar that does not include any contribution from the external heating of the circumstellar envelope by the interstellar radiation field. As a rough estimate, we have used the linear least-square relation between the observed flux at 70 $\mu$m and $L_{\rm int}$, obtained by Dunham et al. (2008) for a set of more than 80 embedded low-luminosity protostars. The $L_{\rm int}$ values are estimated to be $\sim$70-80\% of $L_{\rm bol}$. Thus while $L_{\rm bol}$ may include non-zero contribution from external heating sources, a major fraction of it comes from the central source.

As Fig.~\ref{fig:Lum} indicates, the lowest luminosities of $L_{\rm bol}=0.05~L_\odot$ (J182902.12; $L_{\rm int}$ $\sim$ 0.03 $L_{\sun}$) and $L_{\rm bol}=0.06~L_\odot$ (J182956.67; $L_{\rm int}$ $\sim$ 0.04 $L_{\sun}$) can only be attained by central objects with $M_{\rm obj}$ in the sub-stellar mass regime. Taking into account that $M^{\rm SED}_{\rm{g+d}}$ for these two objects is also low ($\sim$10-12 $M_{Jup}$; Table~\ref{masses}), they will most likely form brown dwarfs. For the following three objects: J182841.87 ($L_{\rm int}$ $\sim$ 0.13 $L_{\sun}$), J182902.84 ($L_{\rm int}$ $\sim$ 0.09 $L_{\sun}$), and J182955.69 ($L_{\rm int}$ $\sim$ 0.15 $L_{\sun}$), the mass of the central object is in the range of $\sim$0.04-0.09 $M_{\sun}$ (Fig.~\ref{fig:Lum}). Considering their low gas+dust mass reservoirs, with $M^{\rm SED}_{\rm{g+d}}$$\sim$0.01-0.02 $M_{\sun}$, they may eventually end up in either the sub-stellar or the very low-mass ($\sim$0.1-0.3 $M_{\sun}$) categories. On the other hand, the remaining two objects, J182855.78 ($L_{\rm int}$ $\sim$ 0.1 $L_{\sun}$) and J182949.57 ($L_{\rm int}$ $\sim$ 0.14 $L_{\sun}$), with $M^{\rm SED}_{\rm{g+d}}$$\sim$0.05-0.2 $M_{\sun}$, can only form very low-mass stars in the long run, even though the current mass of their central objects may still be in the sub-stellar regime. For all of these cases, some of the envelope/disk material may be further photoevaporated or ejected due to jet/outflow activity.

\section{Summary}

We have conducted a multi-wavelength study to identify and characterize proto-brown dwarf candidates in the Serpens Main and Serp/G3-G6 clusters. Our study has revealed four good candidates, as well as a $\sim$20\% fraction of VLM/BD sources mis-identified as embedded YSOs. The lowest luminosity source detected has an observed $L_{\rm bol}$$\sim$0.05 $L_{\sun}$. We have conducted radiative transfer modeling of the observed SEDs and the HCO$^{+}$ (3-2) line profile, and have considered different classification schemes to understand the evolutionary stage of the system. For two candidates, there appears to be only a tenuous dusty envelope, consistent with being Stage I-T/Class Flat sources. Two other sources show more massive envelope, indicative of being Stage 0/I systems. We have set constraints on the mass of the central object using the measured bolometric luminosities and numerical simulations of stellar evolution. Considering the available gas+dust mass reservoir and the current mass of the central source, three of these candidates are likely to evolve into brown dwarfs.

\acknowledgments

We thank the referee for her/his valuable comments that have significantly improved the manuscript. We are grateful to A. Heiderman and N. Evans for providing the HCO$^{+}$ data. BR acknowledges funding from the Marie Sklodowska-Curie Individual Fellowship (Grant Agreement No. 659383). EIV acknowledges support by the Russian Ministry of Education and Science Grant 3.961.2014/K. DH is funded by Deutsche Forschungsgemeinschaft Schwerpunktprogramm (DFG SPP 1385). The James Clerk Maxwell Telescope has historically been operated by the Joint Astronomy Centre on behalf of the Science and Technology Facilities Council of the United Kingdom, the National Research Council of Canada and the Netherlands Organisation for Scientific Research. Additional funds for the construction of SCUBA-2 were provided by the Canada Foundation for Innovation. This research is based on observations using CanariCam at the Gran Telescopio Canarias, a partnership of Spain, Mexico, and the University of Florida, and located at the Spanish Observatorio del Roque de los Muchachos of the Instituto de Astrofisica de Canarias, on the island of La Palma. Based on data obtained from the ESO Science Archive Facility under request number 191891. {\it Herschel} is an ESA space observatory with science instruments provided by European-led Principal Investigator consortia and with important participation from NASA. This work is based [in part] on archival data obtained with the Spitzer Space Telescope, which is operated by the Jet Propulsion Laboratory, California Institute of Technology under a contract with NASA. Support for this work was provided by an award issued by JPL/Caltech. This material is based upon work at the Caltech Submillimeter Observatory, which was operated by the California Institute of Technology under cooperative agreement with the National Science Foundation (AST-0838261).




\begin{deluxetable}{cccccccccccccc}
\tabletypesize{\scriptsize}
\rotate
\tablecaption{Object Positions and Properties of the Bonafide Detections}
\tablewidth{0pt}
\tablehead{
\colhead{Object Name} &  \colhead{RA [J2000]} & \colhead{DEC [J2000]} & \colhead{$L_{\rm bol}$\tablenotemark{a}} & \colhead{$\alpha_{IR}$\tablenotemark{a}} & \colhead{$T_{bol}$\tablenotemark{a}} & HCO$^{+}$ (3-2)\tablenotemark{b} & \colhead{Class\tablenotemark{c}}  & Revised\tablenotemark{d}  \\ 

\colhead{(SSTc2d +)} &  &  & \colhead{($L_{\sun}$)} &  & \colhead{(K)} & Detection &  &  Classification \\
}
\startdata

J182902.12+003120.7 & 18:29:02.1 & +00:31:20.7 & 0.05 &  0.24  & 85 & Y & Flat, I, 0/I  & Flat, I-T \\

J182855.78+002944.8 & 18:28:55.7 & +00:29:44.8 & 0.18 &  1.89 & 54 & Y & 0, 0, 0/I & 0, 0 \\

J182949.57+011706.0 & 18:29:49.5 & +01:17:06.0 & 0.25 &  0.66 & 480 & NA & I, I, -- & 0, I  \\

J182841.87-000321.3 & 18:28:41.8 & -00:03:21.3 & 0.16  & 0.37 & 410 & N & I, Flat, II & Flat, I-T  \\

J182956.67+011239.2 & 18:29:56.6 & +01:12:39.2 & 0.06 &  -0.11 & 690 & Y & Flat, Flat, 0/I & II, II  \\

J182902.84+003009.6 & 18:29:02.8 & +00:30:09.6 & 0.17 &  -0.14 & 460 & NA & Flat, Flat, -- & II, II  \\

J182955.69+011431.6 & 18:29:55.6 & +01:14:31.6 & 0.2 &  -0.23 & 810 & NA & Flat, Flat, -- & II, II  \\

\enddata
\tablenotetext{a}{The estimates for $L_{\rm bol}$, $\alpha_{IR}$, and $T_{bol}$ are from Evans et al. (2009).}
\tablenotetext{b}{Y -- Yes; N -- No; NA -- HCO$^{+}$ (3-2) observations not available.} 
\tablenotetext{c}{The first and second values indicate the classification using the schemes based on $\alpha_{IR}$ and $T_{bol}$, respectively, from the criteria defined in Evans et al. (2009). The third value is the classification using the criteria based on the HCO$^{+}$ line strength from Heiderman \& Evans (2015). }
\tablenotetext{d}{The first value indicates the revised classification based on SED modeling results (Sect.~\ref{model}); the second value is obtained using the ``Stage'' classification scheme.}
\label{properties}
\end{deluxetable}

\begin{deluxetable}{cccccccccccccc}
\tabletypesize{\scriptsize}
\rotate
\tablecaption{Flux Densities of the Bonafide Detections}
\tablewidth{0pt}
\tablehead{
\colhead{Object Name} &  \multicolumn{3}{c}{PACS} & \multicolumn{2}{c}{SCUBA-2} &   \\ 
\colhead{(SSTc2d +)} & \colhead{70 $\mu$m [mJy]} & \colhead{100 $\mu$m [mJy]} & \colhead{160 $\mu$m [mJy]} & \colhead{450 $\mu$m [mJy]} & \colhead{850 $\mu$m [mJy]} & \colhead{1.1 mm [mJy]\tablenotemark{a}} \\
}
\startdata

J182902.12+003120.7 &  330$\pm$28 & 316$\pm$30 & 409$\pm$40 & 79$\pm$8 & 26$\pm$4 & 140$\pm$14 \\

J182855.78+002944.8 &  4621$\pm$370 & 9225$\pm$930 & $<$5220 & 332$\pm$40 & 153$\pm$20 & 970$\pm$97  \\

J182949.57+011706.0 &  834$\pm$84 & 467$\pm$47 & 917$\pm$92 & 1079$\pm$110 & 334$\pm$34 \\

J182841.87-000321.3 &  677$\pm$70 & -- & 918$\pm$90 & 251$\pm$26 & 80$\pm$8 & --  \\

J182956.67+011239.2 &  $<$274 & $<$156 & $<$782 & 251$\pm$25 & 83$\pm$8 & --  \\

J182902.84+003009.6 & 268$\pm$26 & 273$\pm$28 & $<$508 & 136$\pm$13 & 43$\pm$4 & 180$\pm$18 \\

J182955.69+011431.6 &  $<$664 & $<$347 & $<$842 & 326$\pm$32 & 150$\pm$15 & -- \\

\enddata
\tablenotetext{a}{The 1.1 mm flux densities are from Enoch et al. (2009).}
\label{phot}
\end{deluxetable}

\begin{deluxetable}{lccccccccccccc}
\tabletypesize{\scriptsize}
\rotate
\tablecaption{SED Model Fit Parameters}
\tablewidth{0pt}
\tablehead{
\colhead{Object Name} & \colhead{$M_{\rm env}$\tablenotemark{a}} & \colhead{$M_{\rm disk}$\tablenotemark{a}} & \colhead{$\dot{M}_{env}$} & \colhead{$R_{\rm env,max}$} & \colhead{$R_{\rm d,max}$} & \colhead{$\theta_{\rm in}$}  \\

\colhead{(SSTc2d +)} & \colhead{($M_{\rm Jup}$)} & \colhead{($M_{\rm Jup}$)} & \colhead{($M_{\sun}$ yr$^{-1}$)}  & \colhead{(au)} & \colhead{(au)} & \colhead{($\degr$)} & \\
}
\startdata

J182902.12+003120.7 & 8$\pm$3 & 2$\pm$1 & (0.3$\pm$0.1)$\times$10$^{-6}$  & 1510$\pm$100 & 102$\pm$30 & 80$\pm$10  \\
J182855.78+002944.8 & 216$\pm$10 & 7$\pm$2 & (2$\pm$1)$\times$10$^{-5}$ & 1700$\pm$100 & 47$\pm$5 & 32$\pm$10  \\
J182949.57+011706.0 & 33$\pm$10 & 17$\pm$8 &(2$\pm$1)$\times$10$^{-6}$  & 2230$\pm$500 & 10$\pm$5 & 40$\pm$10  \\
J182841.87-000321.3 & 13$\pm$6 & 6$\pm$1 & (0.4$\pm$0.2)$\times$10$^{-6}$ & 1420$\pm$400 & 46$\pm$6 & 40$\pm$10  \\
J182956.67+011239.2 & 0.4$\pm$0.3 & 12$\pm$4 & (2$\pm$1)$\times$10$^{-9}$ & -- & 73$\pm$20 & 80$\pm$10  \\
J182902.84+003009.6 & $<$10$^{-6}$ & 11$\pm$3 & --  & -- & 11$\pm$3 & 75$\pm$10  \\
J182955.69+011431.6 & 0.1$\pm$0.05 & 18$\pm$5 & (0.2$\pm$0.1)$\times$10$^{-9}$ & -- & 90$\pm$20 & 75$\pm$10   \\

\enddata
\tablenotetext{a}{$M_{\rm env}$ and $M_{\rm disk}$ are the (gas+dust) masses.}
\label{SEDpars}
\end{deluxetable}

\begin{deluxetable}{ccccccccccl}
\tabletypesize{\scriptsize}
\rotate
\tablecaption{Total Masses}
\tablewidth{0pt}
\tablehead{
\colhead{Object Name} & \colhead{$M_{g+d}^{SED}$} & \colhead{$M_{g+d}^{850}$} & \colhead{$T_{dust}$} & \colhead{$S_{model}$} & \colhead{$S_{obs}$} \\ 

\colhead{(SSTc2d +)} &  \colhead{($M_{Jup}$)} & \colhead{($M_{Jup}$)} & \colhead{(K)} & \colhead{($\arcsec$)} & \colhead{($\arcsec$)}  \\

}
\startdata

J182902.12+003120.7 &10$\pm$3 & 31$\pm$4 & 15 & 6$\pm$0.4 & 16$\pm$4  \\

J182855.78+002944.8 & 223$\pm$10 & 178$\pm$24 & 9 & 7$\pm$0.4 & 16$\pm$4  \\

J182949.57+011706.0 & 50$\pm$13 & 390$\pm$40 & 35 & 9$\pm$2 & 8$\pm$4  \\

J182841.87-000321.3 &19$\pm$6 & 93$\pm$10 & 40 & 6$\pm$1.5 & 16$\pm$4   \\

J182956.67+011239.2 & 12$\pm$4 & 97$\pm$9 & 40 & 0.3$\pm$0.08 & 8$\pm$4  \\

J182902.84+003009.6 & 11$\pm$3 & 50$\pm$5 & 25 & 0.05$\pm$0.01 & 12$\pm$4   \\

J182955.69+011431.6 & 18$\pm$5 & 174$\pm$19 & 40 & 0.4$\pm$0.08 & 16$\pm$4  \\  
   
\enddata
\label{masses}
\end{deluxetable}

\begin{deluxetable}{cccccccccccccc}
\tabletypesize{\scriptsize}
\rotate
\tablecaption{Marginal Detections in the SCUBA-2 Maps}
\tablewidth{0pt}
\tablehead{
\colhead{Object Name} &  \colhead{RA [J2000]} & \colhead{DEC [J2000]} & \colhead{$\alpha_{IR}$\tablenotemark{a}} & \colhead{$L_{\rm bol}$} & \colhead{$T_{bol}$} & \colhead{Class\tablenotemark{b}} &  \multicolumn{2}{c}{SCUBA-2\tablenotemark{c}}    \\ 
\colhead{(SSTc2d +)} &  &  &  & \colhead{($L_{\sun}$)} & \colhead{(K)} & &  \colhead{450 $\mu$m [mJy]} & \colhead{850 $\mu$m [mJy]}  \\
}
\startdata

J183002.08+011359.0 & 18:30:02.08 & +01:13:59.0 & 0.36 & 0.09 & 510 & I, I & $<$54 & $<$96 \\
J182959.38+011041.1 & 18:29:59.38 & +01:10:41.1 & 0.51 & 0.008 & 440 & I, I & $<$17 & $<$57 \\
J183018.17+011416.9 & 18:30:18.17 & +01:14:16.9 & -0.29 & 0.13 & 1200 & Flat, II & $<$15 & $<$41 \\
J182952.21+011559.1 & 18:29:52.21 & +01:15:59.1 & 1.87 & 0.024 & 230 & I, I & $<$273 & $<$137 \\

\enddata
\tablenotetext{a}{The estimates for $\alpha_{IR}$, Class, $L_{\rm bol}$, and $T_{bol}$ are from Evans et al. (2009).}
\tablenotetext{b}{First value is based on $\alpha_{IR}$, second value is based on the $T_{bol}$ classification ranges, from Evans et al. (2009).}
\tablenotetext{c}{Upper limits.}
\label{marg}
\end{deluxetable}

 \begin{figure*}
  \centering              
     \includegraphics[width=1.5in]{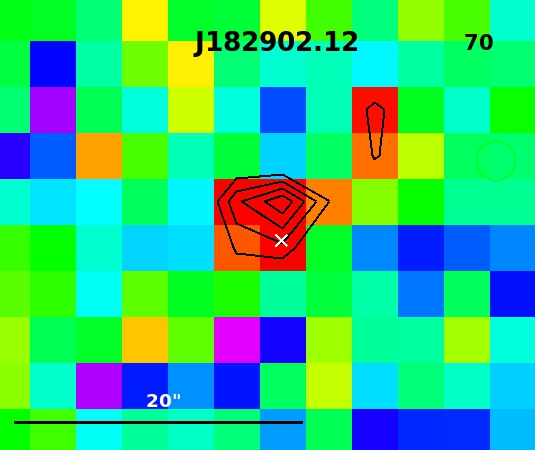}  
     \includegraphics[width=1.5in]{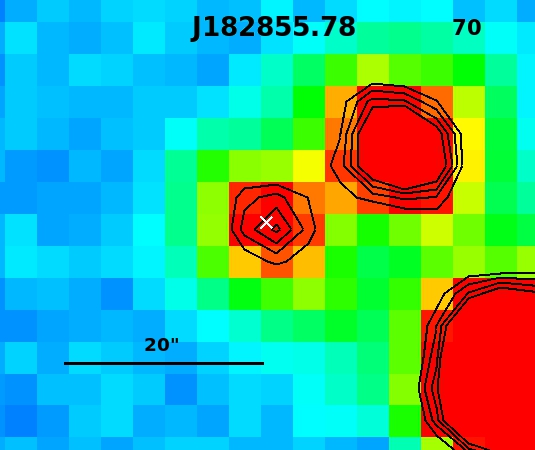}   
     \includegraphics[width=1.5in]{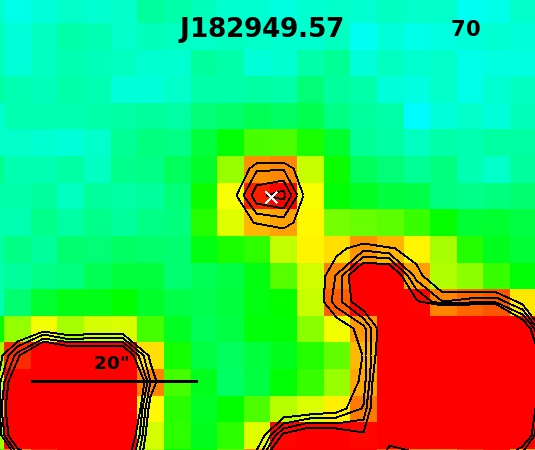}         \\        
        \vspace{0.05in}  
     \includegraphics[width=1.5in]{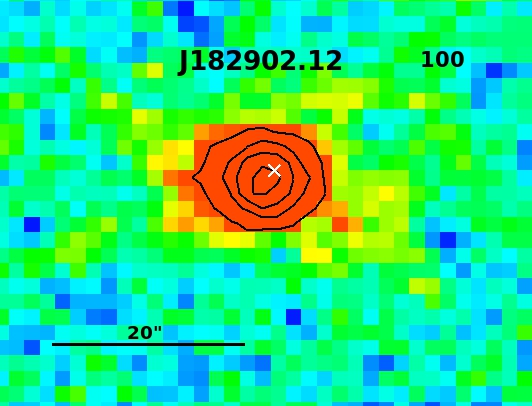}  
     \includegraphics[width=1.5in]{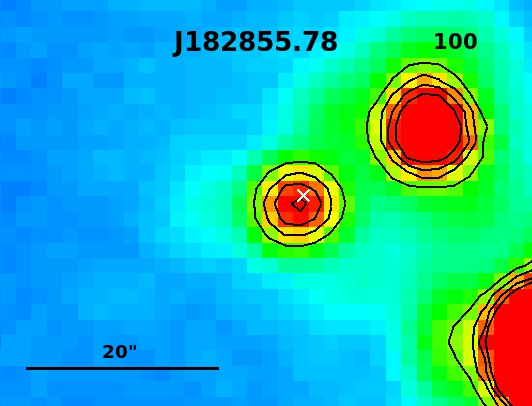}    
     \includegraphics[width=1.5in]{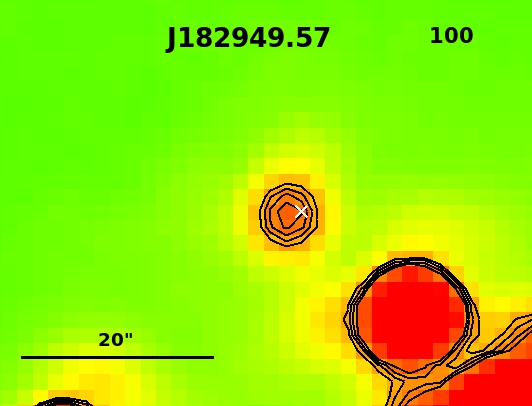}          \\          
        \vspace{0.05in}       
     \includegraphics[width=1.5in]{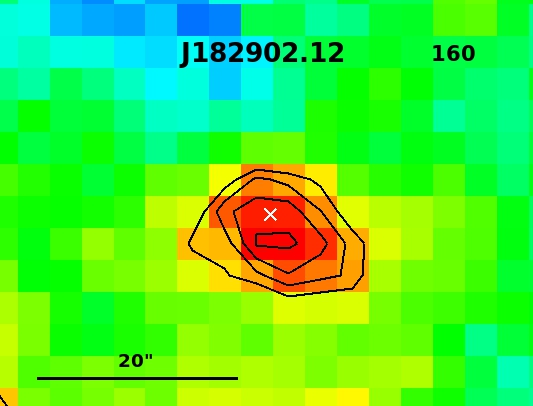}  
     \includegraphics[width=1.5in]{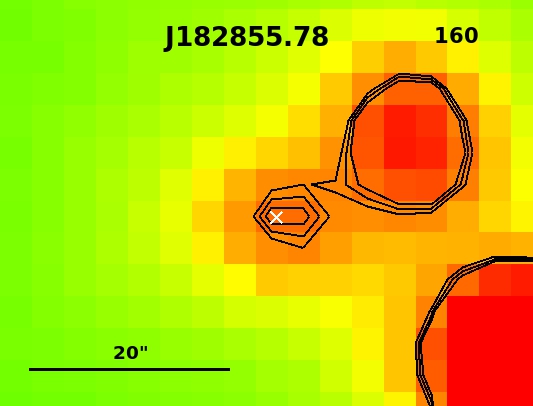}    
     \includegraphics[width=1.5in]{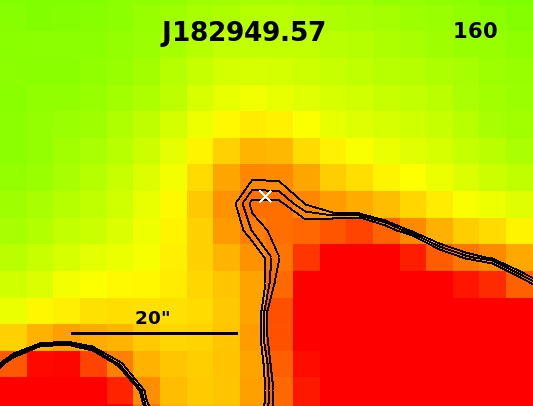}          \\            
        \vspace{0.05in}       
     \includegraphics[width=1.5in]{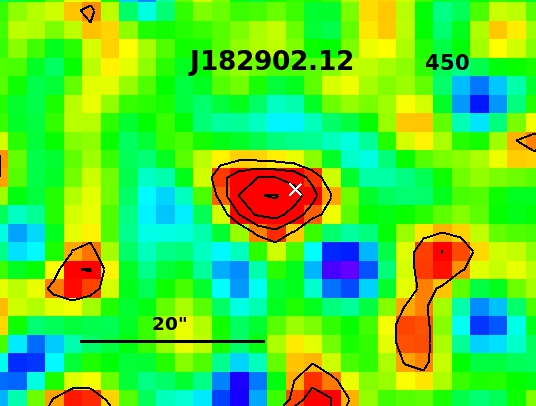}  
     \includegraphics[width=1.5in]{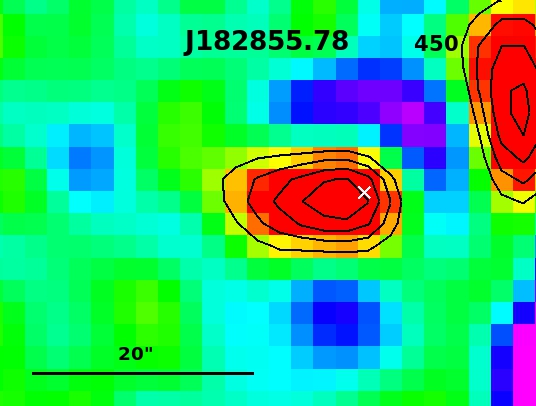}    
     \includegraphics[width=1.5in]{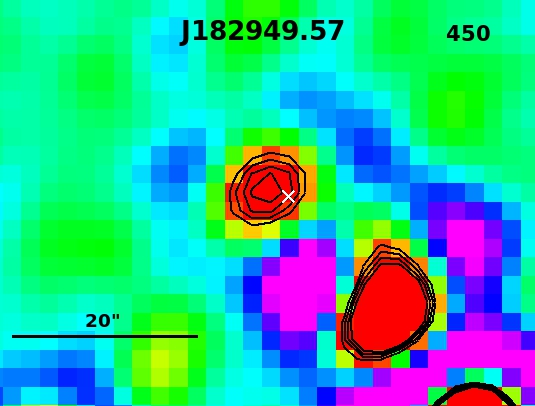}        \\           
        \vspace{0.05in}  
     \includegraphics[width=1.5in]{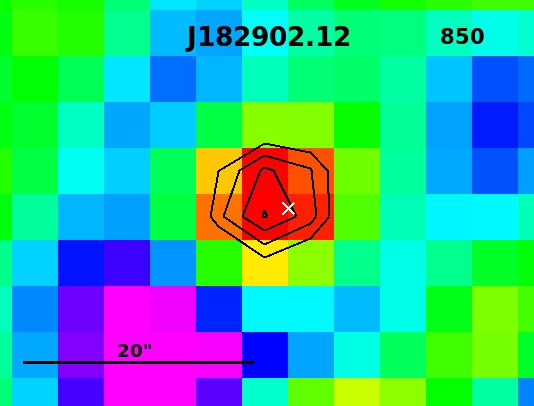}  
     \includegraphics[width=1.5in]{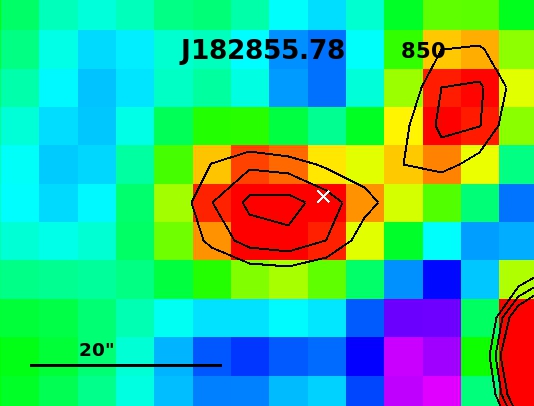}           
     \includegraphics[width=1.5in]{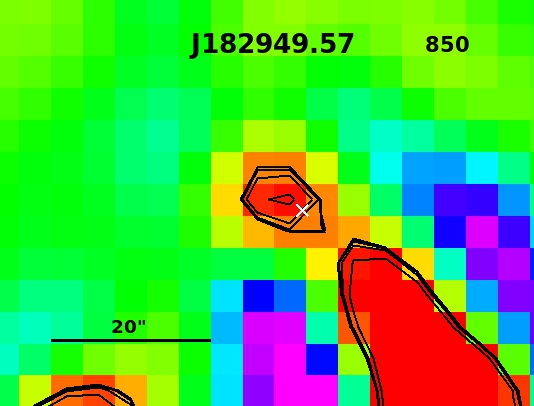}     \\ 
     \includegraphics[width=4.6in]{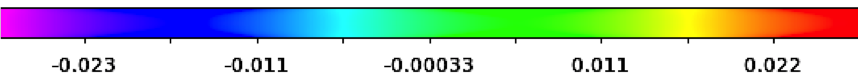}      
     
           \caption{Herschel and SCUBA-2 images for the bonafide sources. Target position is marked by a cross. From top to bottom, the images are at 70 $\mu$m, 100 $\mu$m, 160 $\mu$m, 450 $\mu$m, and 850 $\mu$m. The sources from left to right are J182902.12, J182855.78, and J182949.57. The representative color scale at the bottom shows the intensity in units of Jy beam$^{-1}$. The contours are given in steps of one from 1$\times$ to 5$\times$ the peak intensity. The pixel size in the PACS 70 $\mu$m and 100 $\mu$m bands is 3.2$\arcsec$, and 6.4$\arcsec$ in the 160 $\mu$m band. The default map pixels are 2$\arcsec$ and 4$\arcsec$ at 450 $\mu$m and 850 $\mu$m, respectively. The spatial scale is shown at the bottom right. North is up, east is to the left. }
          \label{imgs-bf}
  \end{figure*}

 \begin{figure}
  \centering              
     \includegraphics[width=3in]{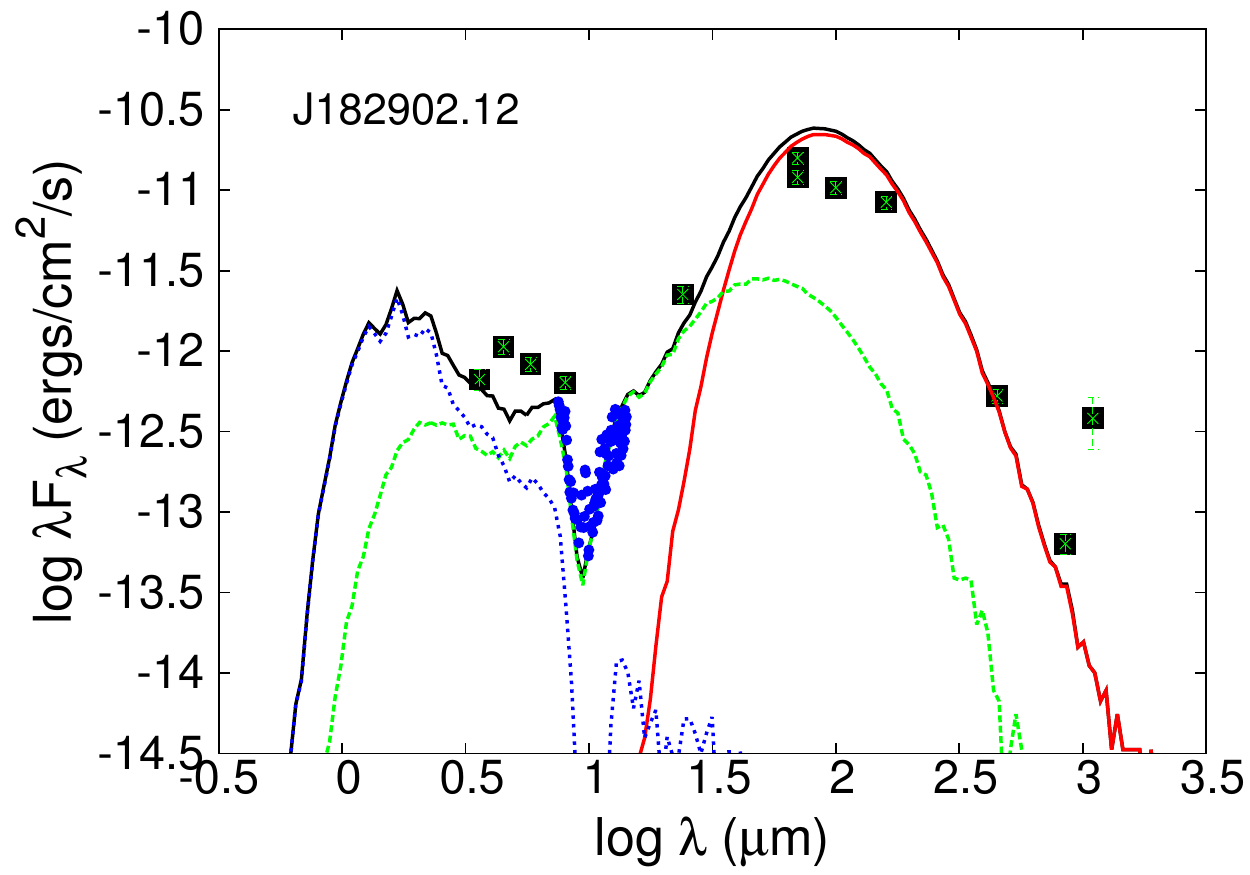}
     \includegraphics[width=3in]{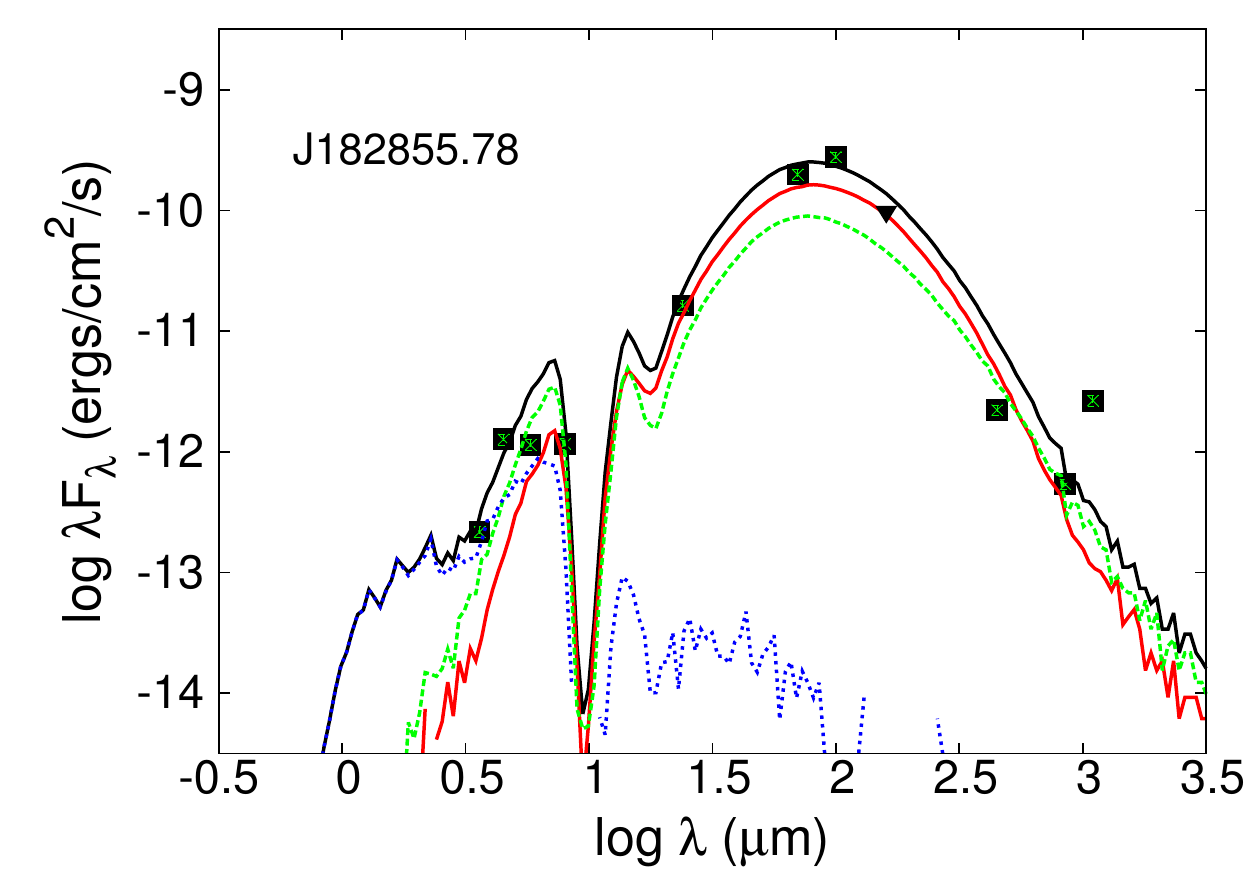}    
     \includegraphics[width=3in]{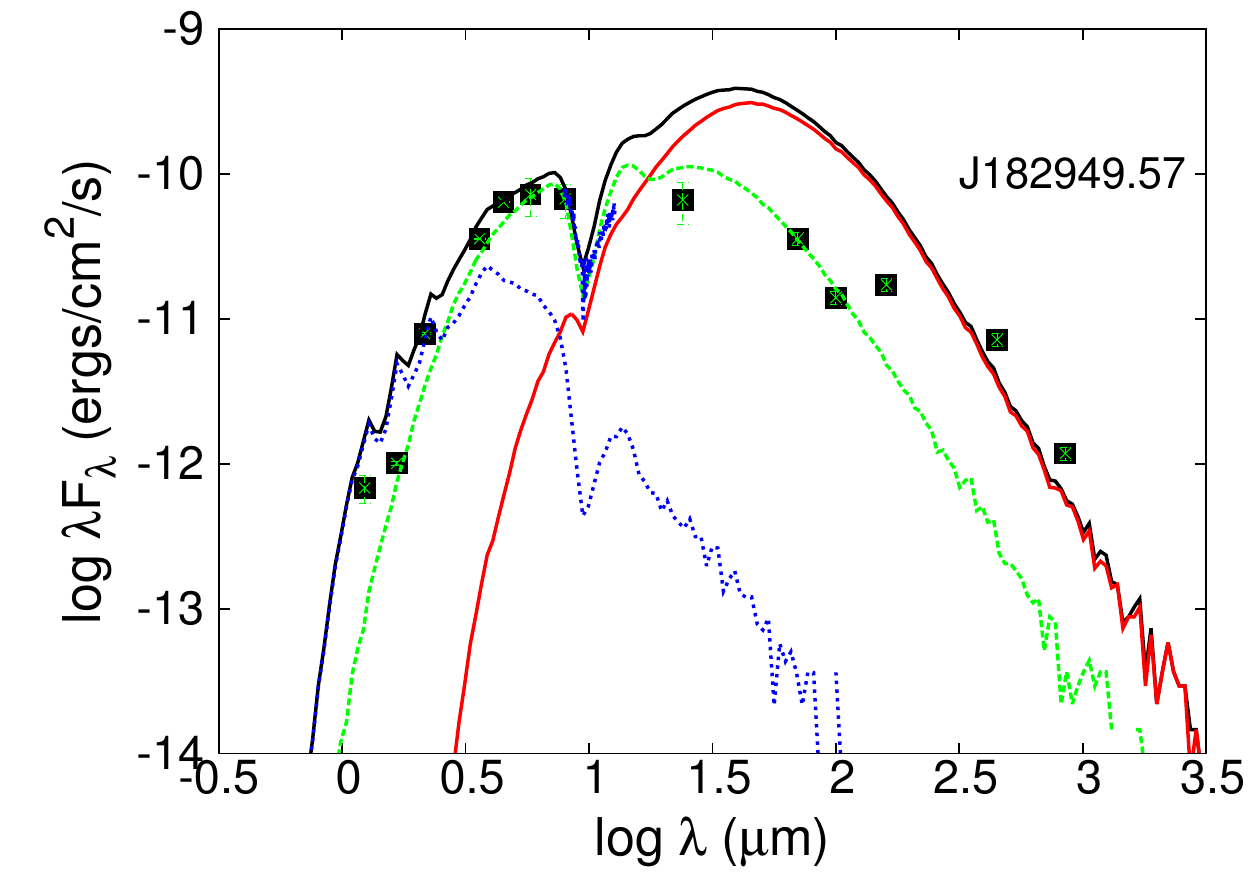}  
     \caption{The SEDs with model fits for the bonafide sources. Red, green, and blue lines indicate the individual contribution from the envelope, disk, and stellar components, respectively.  }
     \label{seds-bf}
  \end{figure}

 \begin{figure*}
  \centering              
     \includegraphics[width=1.4in]{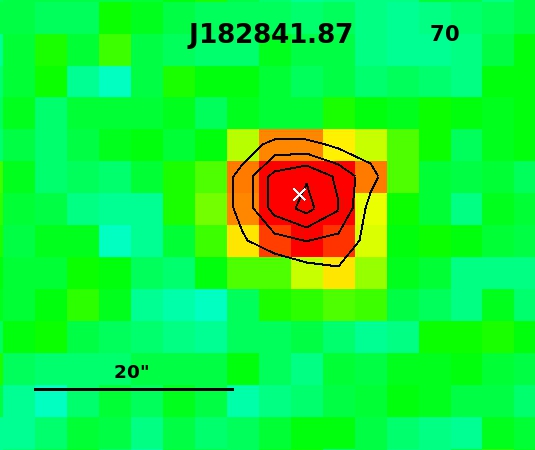}  
     \includegraphics[width=1.4in]{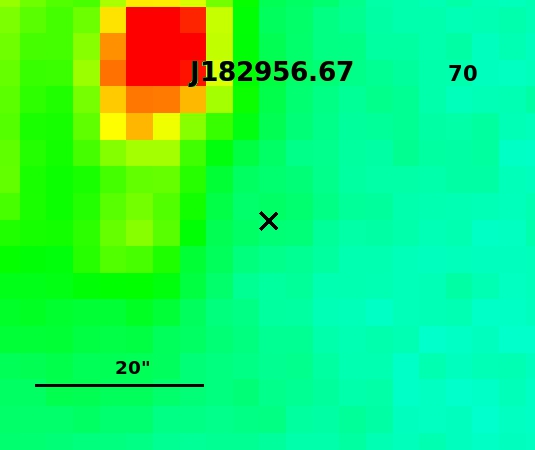}  
     \includegraphics[width=1.4in]{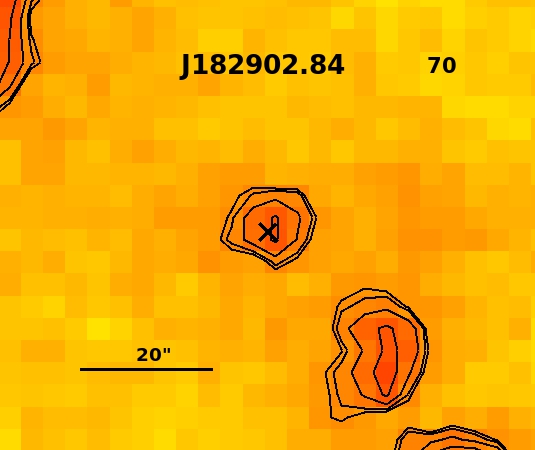}     
     \includegraphics[width=1.4in]{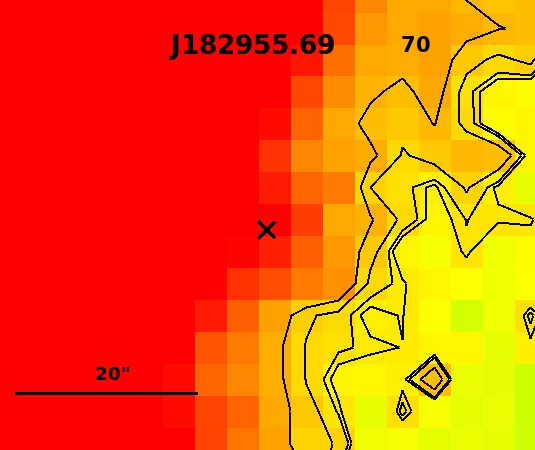}       \\       
     \hspace{1.4in}                 
     \includegraphics[width=1.4in]{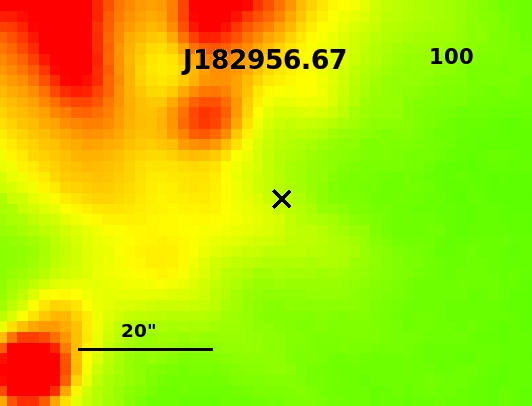}  
     \includegraphics[width=1.4in]{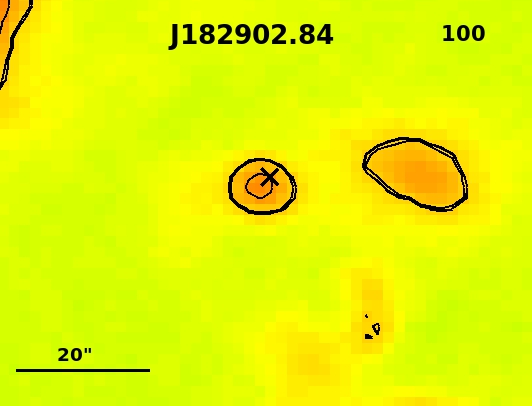}  
     \includegraphics[width=1.4in]{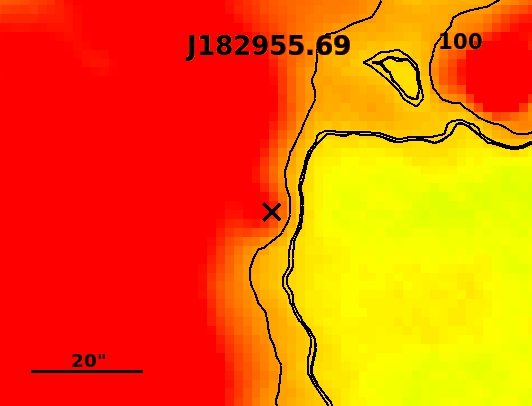}     \\
     \includegraphics[width=1.4in]{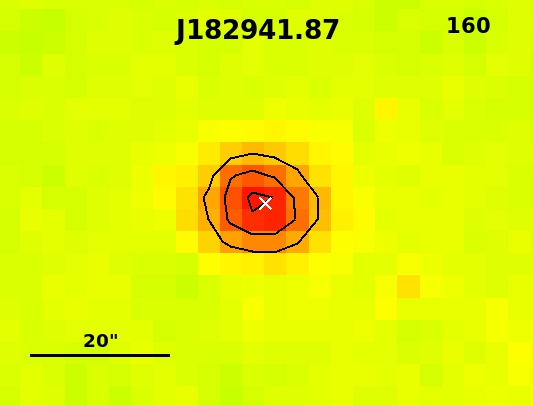}                     
     \includegraphics[width=1.4in]{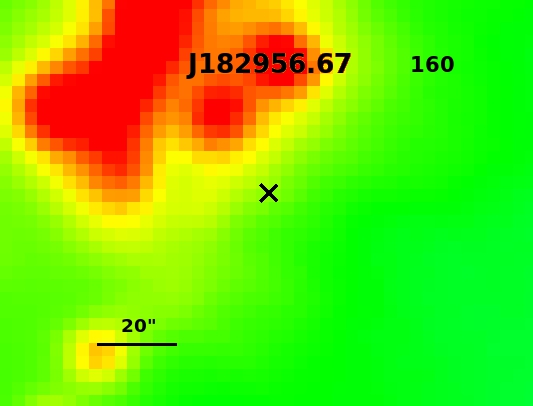}  
     \includegraphics[width=1.4in]{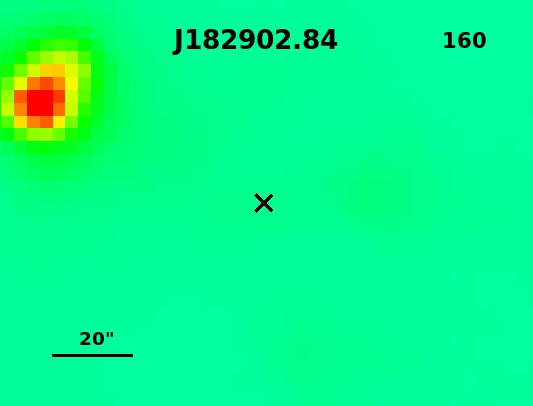}  
     \includegraphics[width=1.4in]{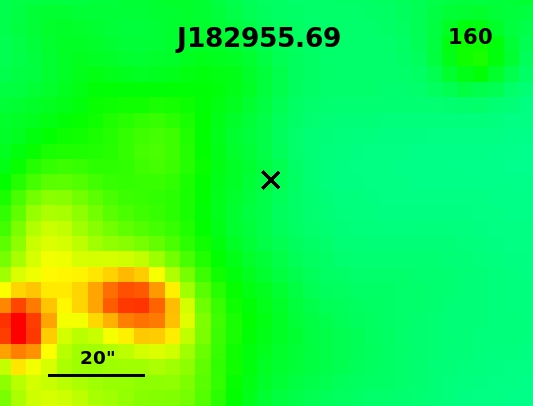}      \\
     \includegraphics[width=1.4in]{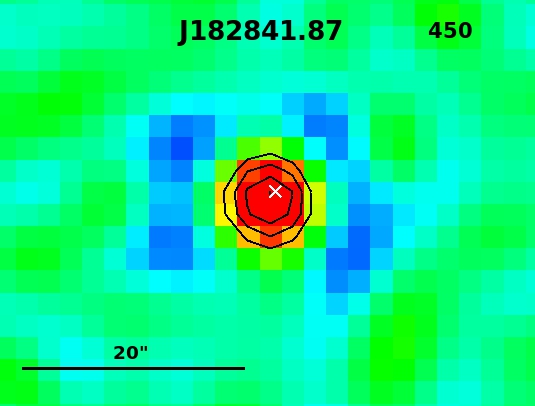}                 
     \includegraphics[width=1.4in]{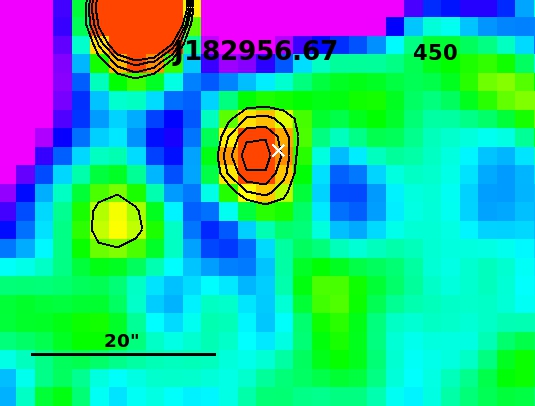}  
     \includegraphics[width=1.4in]{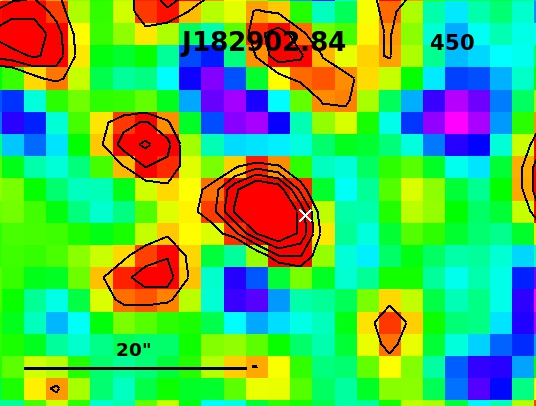}    
     \includegraphics[width=1.4in]{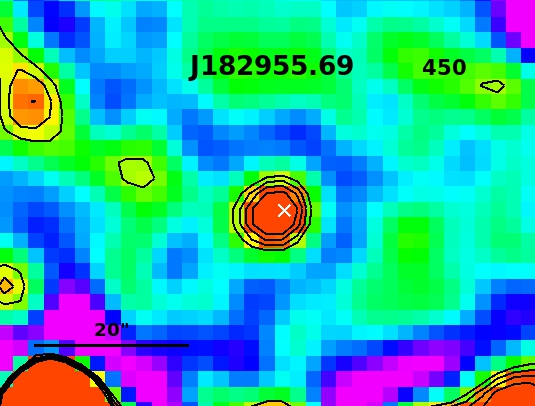}         \\             
     \includegraphics[width=1.4in]{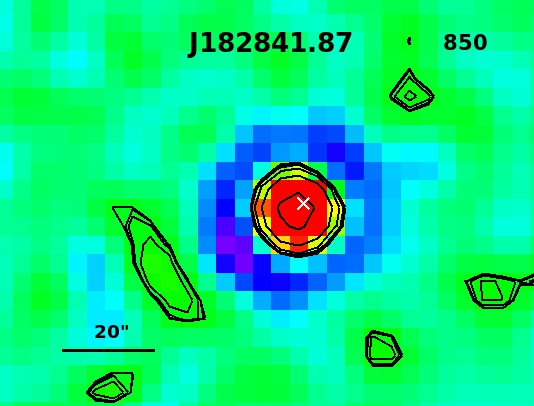} 
     \includegraphics[width=1.4in]{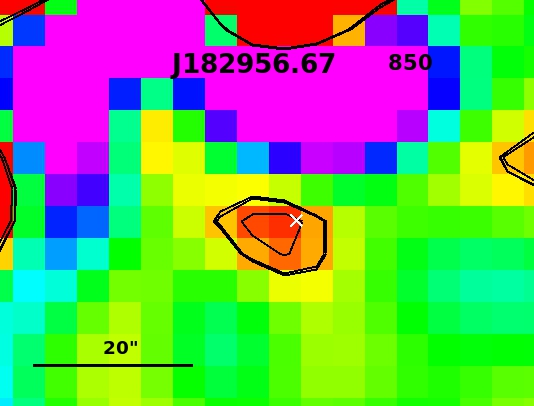}  
     \includegraphics[width=1.4in]{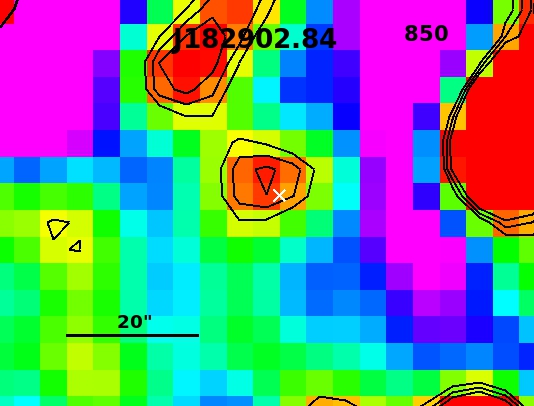}          
     \includegraphics[width=1.4in]{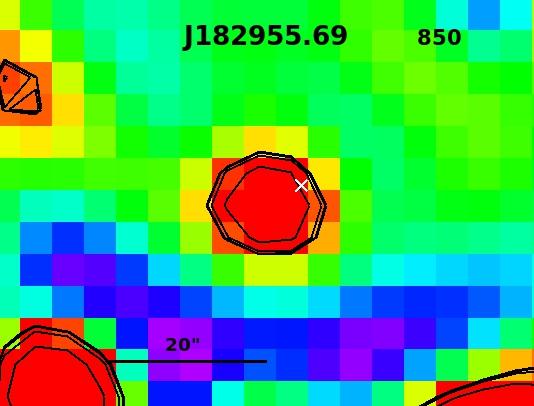}   
     \includegraphics[width=5.5in]{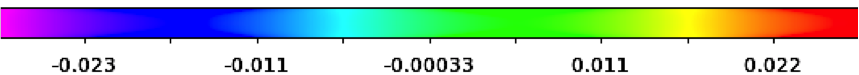}      
                  
     \caption{Herschel and SCUBA-2 images for the mis-identified sources. From top to bottom, the images are at 70 $\mu$m, 100 $\mu$m, 160 $\mu$m, 450 $\mu$m, and 850 $\mu$m. The sources from left to right are J182841.87, J182956.67, J182902.84, and J182955.69. The representative color scale at the bottom shows the intensity in units of Jy beam$^{-1}$. The contours are given in steps of one from 1$\times$ to 3$\times$ the peak intensity. The pixel size in the PACS 70 $\mu$m and 100 $\mu$m bands is 3.2$\arcsec$, and 6.4$\arcsec$ in the 160 $\mu$m band. The default map pixels are 2$\arcsec$ and 4$\arcsec$ at 450 $\mu$m and 850 $\mu$m, respectively. The spatial scale is shown at the bottom right. North is up, east is to the left.   }
     \label{imgs-mis}
  \end{figure*}

 \begin{figure}
  \centering         
     \includegraphics[width=3in]{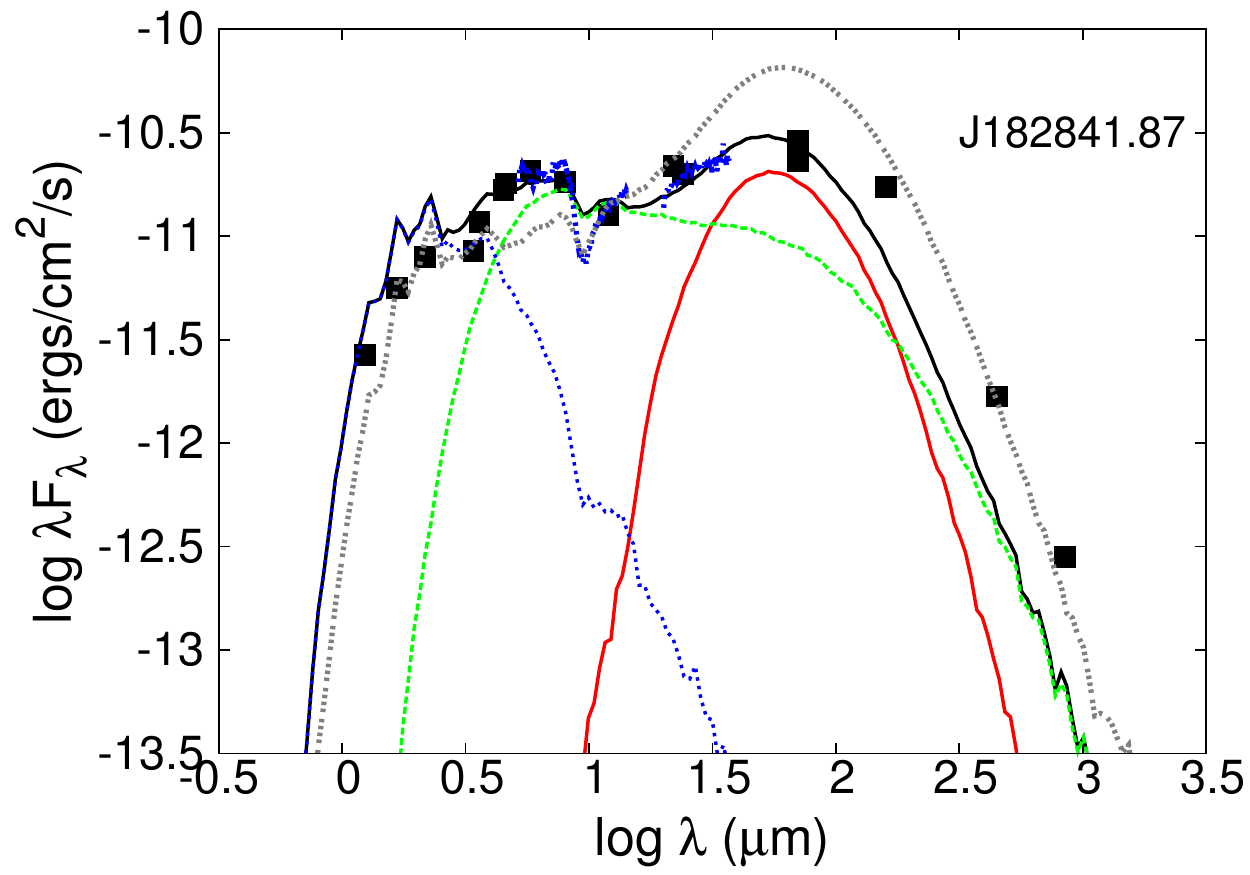}  
     \includegraphics[width=3in]{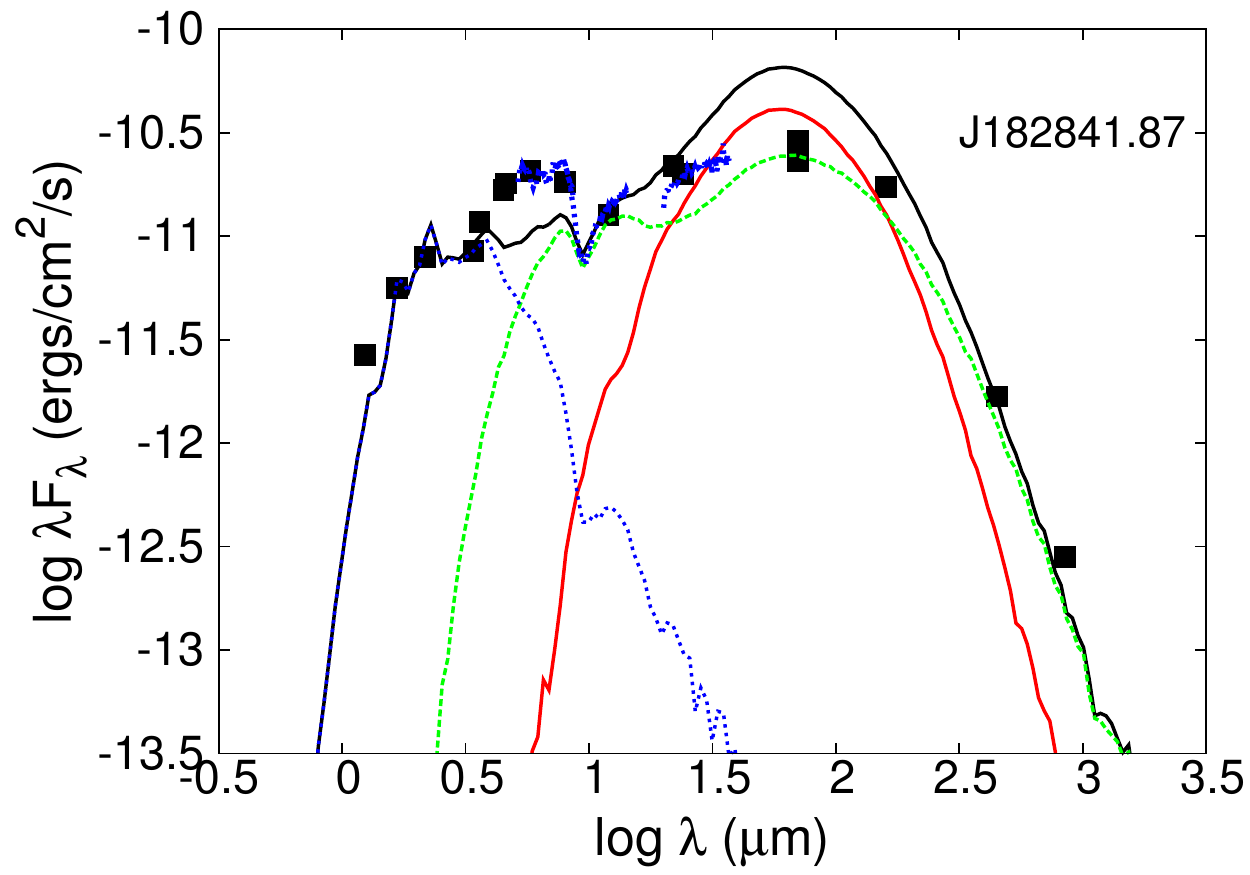}  \\
     \includegraphics[width=3in]{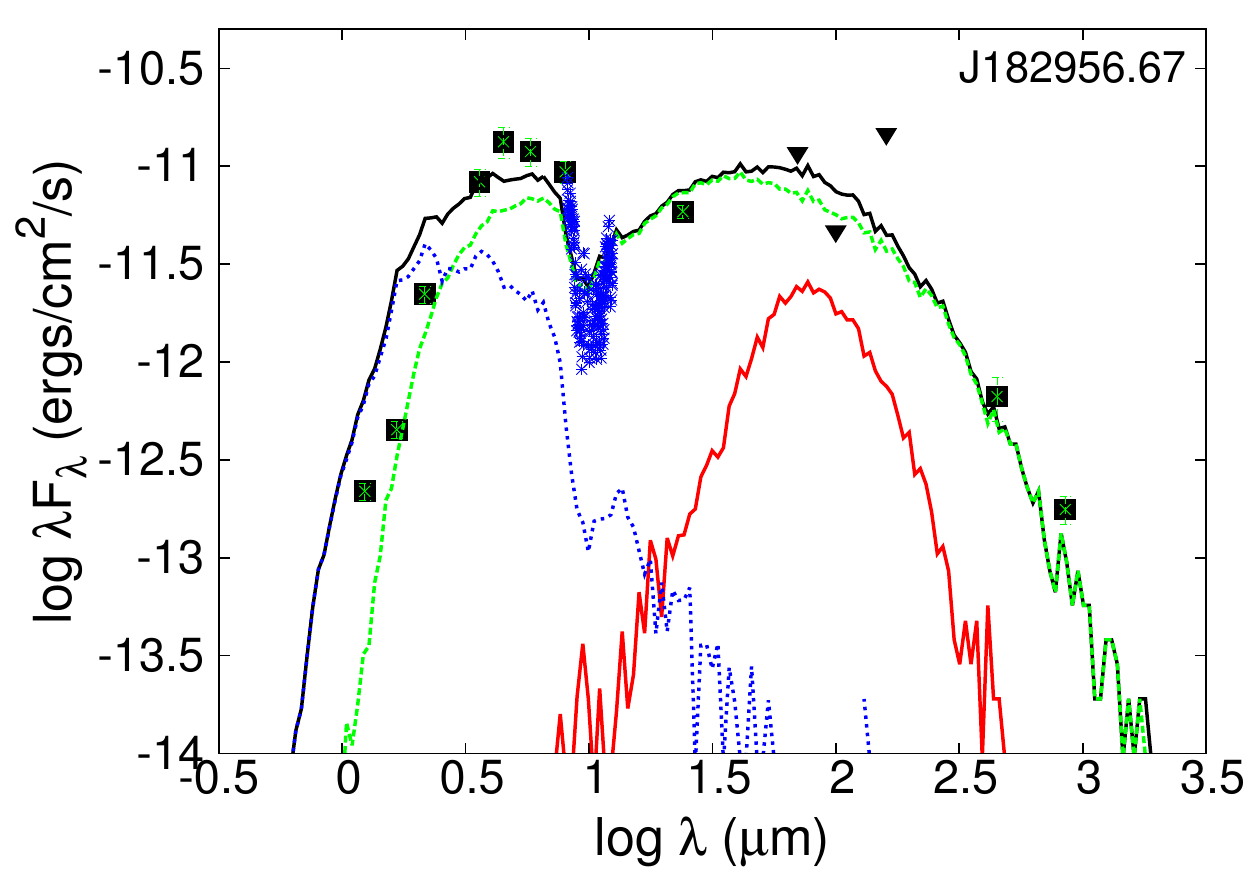} 
     \includegraphics[width=3in]{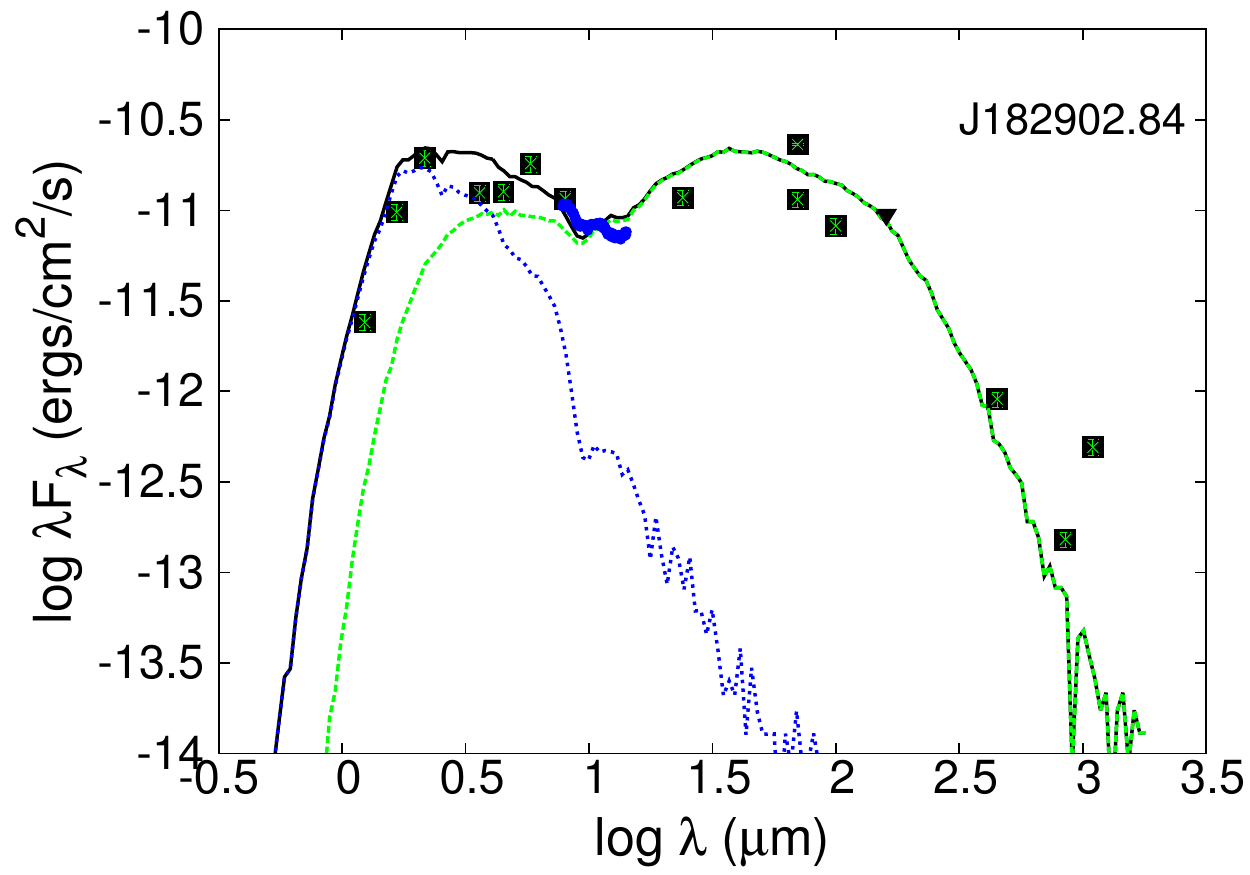}         
     \includegraphics[width=3in]{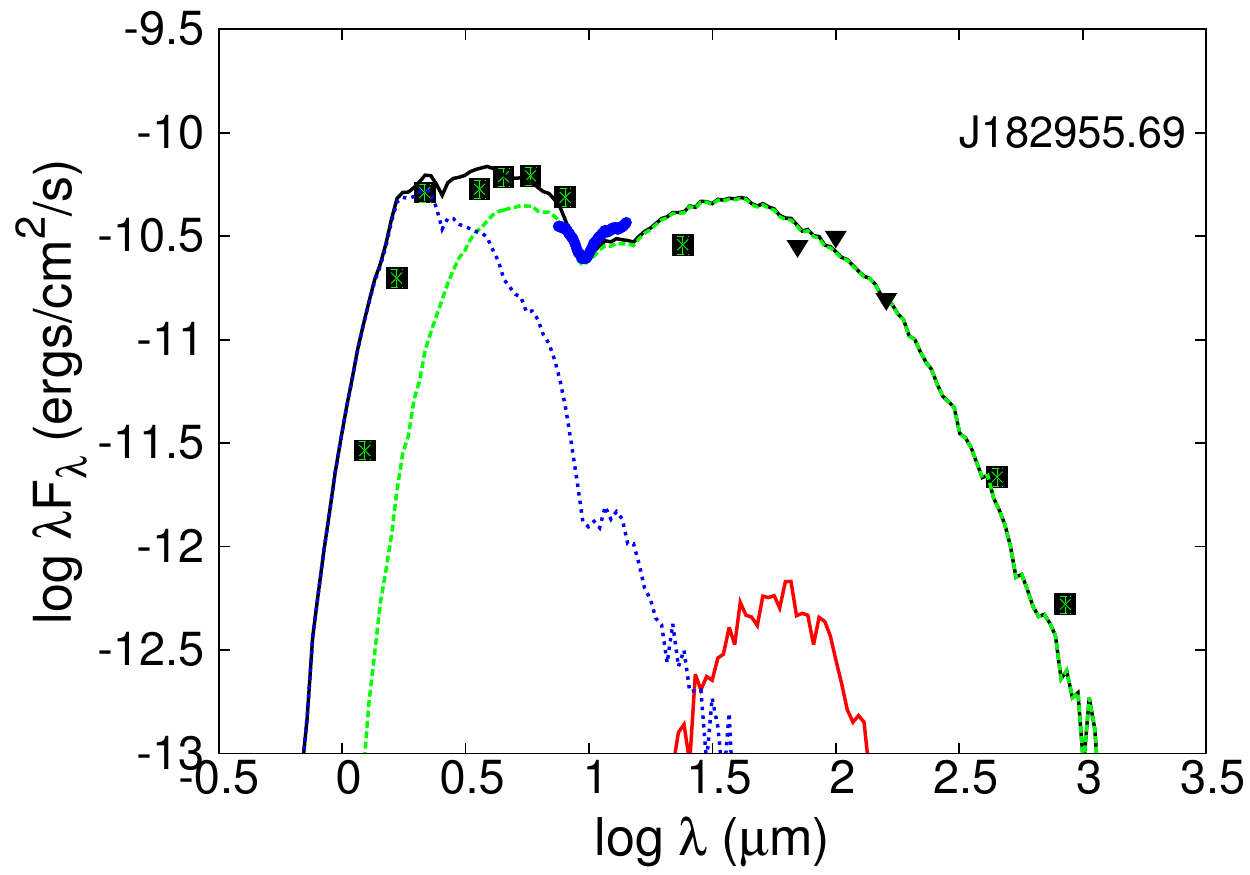}       
     \caption{The SEDs with model fits for the mis-identified sources. Top panel shows the two possible fits for J182841.87. Red, green, and blue lines indicate the individual contribution from the envelope, disk, and stellar components, respectively.}
     \label{seds-mis}
  \end{figure}

\newpage

\section{Appendix}
\label{appendix}

\subsection{Marginal Detections}  
\label{marginal}

There are three sources that are detected at a marginal ($\sim$2-$\sigma$) level in the SCUBA-2 850\,$\mu$m band, but are undetected at 450\,$\mu$m as well as in the PACS bands. These sources are listed in Table~\ref{marg} with the SCUBA-2 photometry, which should be considered as upper limits. None of these objects have prior detections reported at 1.1 mm or in the MIPS 70 $\mu$m band. There are no HCO$^{+}$ (3-2) observations available for these sources. For all of these cases, the best-fits from modeling the SEDs including the PACS and 450\,$\mu$m upper limits indicate the presence of an edge-on disk of a very low-mass of $\sim$1-4 $M_{Jup}$. The spectral slope of -0.29 for J183018.17+011416.9 is very close to the Class Flat/II boundary, so this is likely a disk only source. It has been classified as M2.9$\pm$0.6 YSO (Gorlova et al. 2010). 

Among the Class I sources, SSTc2d J183002.08+011359.0 lies in a nebulous region, and J182959.38+011041.1 lies in the filament seen in the south-west sub-cluster. It is likely that the excess emission seen in the mid-infrared for these sources is due to the surrounding cold material rather than the source itself, resulting in the steep rise in the SED observed between 3.6$\mu$m and 4.5$\mu$m and a high value for the spectral slope. SSTc2d J182959.38+011041.1 also has a very low observed $L_{\rm bol}$ of 0.008 $L_{\sun}$, while the lowest luminosity source detected in our SCUBA-2 observations has $L_{\rm bol}$ $\sim$ 0.05 $L_{\sun}$. In contrast, the marginal detection of the relatively more luminous Class I source SSTc2d J183002.08+011359.0 is perplexing. Deeper scans can help further characterize these sources. 


\subsection{Unresolved/Confused Sources}
\label{unresolved}

We have found four sources that are unresolved from a nearby bright YSO in the SCUBA-2 maps. The lowest luminosity case amongst these is SSTc2d J182909.05+003128.0 ($L_{\rm bol}$ = 0.0078 $L_{\sun}$) that lies at a $\sim$2$\arcsec$ separation from the protostar SSTc2d 182909.07+003132.4 ($L_{\rm bol}$ = 1.1 $L_{\sun}$). The two objects are also unresolved in the PACS bands. While there are separate HCO$^{+}$ (3-2) pointings for the two sources, the observed profiles are very similar. The HCO$^{+}$ profiles for both sources show weak emission centered at the systemic velocity. Considering the large beamsize for these molecular line observations, the two pointings must be tracing the composite source. For another very low-luminosity case of SSTc2d J182952.21+011559.1 ($L_{\rm bol}$ = 0.024 $L_{\sun}$), there is no object seen at the source location in the SCUBA-2 450\,$\mu$m map. This object is at a $\sim$13$\arcsec$ separation to the bright protostar SSTc2d J182952.08+011547.8, and is within the beam size of $\sim$14.5$\arcsec$ in the 850\,$\mu$m band. This is a case of both a non-detection and an unresolved object due to the poor sensitivity and angular resolution in the SCUBA-2 maps.

Among the more luminous ($L_{\rm bol}$ $\sim$ 0.1 $L_{\sun}$) Class Flat sources, SSTc2d J182957.66+011304.6 is in close proximity ($\leq$8$\arcsec$) to the pre-stellar core JCMTSF J182956.6+011309, while the faint object SSTc2d J182949.69+011456.8 is missed in the bright halo of the pre-stellar core JCMTSF J182949.8+011515. The HCO$^{+}$ line for both sources shows strong emission with blue-dominated axisymmetric profile, indicative of an infalling envelope. The strong emission, however, is caused by the massive sources within the 25$\arcsec$ beamsize of the HCO$^{+}$ observations. Due to the HCO$^{+}$ detection, all of these faint unresolved/confused sources have been categorized as Stage 0/I embedded objects (Heidermann \& Evans 2015), which is incorrect. The classification for these objects is questionable as the sharp rise in the SED could be contaminated by the nearby protostar. 

An additional source SSTc2d J183000.30+010944.7 ($\alpha_{IR}$ = -0.12, $L_{\rm bol}$=0.09$L_{\sun}$) is undetected in our continuum observations. This object is located in the filament at the very south of the south-west sub-cluster. There is no point source seen in the SCUBA-2 and PACS maps at the object location except confusion noise, which may have caused the Class I SED shape and the high $L_{\rm bol}$ for this object. The actual source might be fainter than the observed luminosity. The HCO$^{+}$ line shows a symmetric profile centered at the systemic velocity. The $\alpha_{IR}$ for this source is closer to the Class Flat/Class II threshold, and this might be a Class II object instead of the Stage 0/I classification determined from HCO$^{+}$ detection. Higher resolution observations are required to resolve these source and confirm their evolutionary stage.

\subsection{Consequences of uncertainties}
\label{caveats}

For three sources in our study, J182902.12, J182855.78, and J182902.84, there is 1.1 mm photometry available from CSO Bolocam observations presented in Enoch et al. (2009). For all three cases, the 1.1 mm flux density is higher than the 850\,$\mu$m photometry, and is notably offset from the best model-fit to the SED (Figs.~\ref{seds-bf};~\ref{seds-mis}). The beamsize for CSO Bolocam is almost twice that of SCUBA-2 850\,$\mu$m band ($\sim$14.5$\arcsec$). As can be seen in the 850 $\mu$m images, there is a bright source about 18$\arcsec$ away from the target J182855.78, while J182902.84 and J182902.12 lie in regions where diffuse nebulosity is seen about 20-25$\arcsec$ from the target position (Figs.~\ref{imgs-bf};~\ref{imgs-mis}). It appears that for all three cases, the flux within the Bolocam 30$\arcsec$ beam is likely enhanced due to the contribution from another source within the beam, or a higher contribution from the surrounding material, whereas there is less contribution from these contaminants in the 14$\arcsec$ SCUBA-2 beamsize. 

As discussed in Sect.~\ref{submm-obs}, we have applied the masking technique to the SCUBA-2 maps, which can provide the optimum source contribution, particularly for such faint objects. The 850\,$\mu$m photometry is thus more reliable, since a similar masking technique has not been applied to the published 1.1 mm photometry. We have opted not to use the 1.1 mm point in fitting the SEDs; any model that provides a good fit to this point misses the sub-millimeter and far-infrared points, and is not a good fit to the overall SED. Nevertheless, we have attempted to fit the 1.1 mm point for the case of J182902.12, which shows the largest offset in the 1.1 mm photometry (Fig.~\ref{seds-bf}). The model that provides a good fit to the 1.1 mm point shows more than half an order of magnitude higher fluxes at far-infrared and sub-millimeter wavelengths than the observed photometry, and is for an envelope mass of $\sim$14 $M_{Jup}$ and a disk mass of $\sim$3 $M_{Jup}$, higher than the mass estimates obtained from the best-fit (Table~\ref{SEDpars}). Likewise, the envelope+disk mass for the J182855.78 system from a model that fits the 1.1 mm point is $\sim$260 $M_{Jup}$, while a similar fit for J182902.84 is for a disk mass of $\sim$24 $M_{Jup}$. As discussed in Sect.~\ref{totalmass}, $M_{obj}$ for J182902.12 is in the sub-stellar regime, and with an addition of envelope + disk mass of $\sim$18 $M_{Jup}$, this system is still likely to reach a final mass below the sub-stellar limit, while J182855.78 and J182902.84 will likely evolve into very low-mass stars. However, the reduced-$\chi^{2}$ value for such a fit is $>$5, much poorer than the fit obtained excluding the 1.1 mm point. The alternative total mass obtained by fitting the 1.1 mm photometry is thus not reliable. Furthermore, in order to compare the total masses and physical properties for all sources in this study (Sect.~\ref{discussion}), it is important to be consistent and use the measurements obtained by applying the same methodologies on the same set of data, rather than using the masses obtained from the 1.1 mm photometry for just three sources. 

Finally, the distance to Serpens is still uncertain, as discussed in Sect.~\ref{intro}. While nearly all distance estimates are within $\sim$260-310 pc, the Dzib et al. (2010) measurement of 415 pc would imply that the $L_{bol}$ estimates for our targets would be higher by a factor of $\sim$2.5. Based on the arguments provided in Sect.~\ref{totalmass}, this would place the lowest luminosity sources J182902.12 and J182956.67 in the very low-mass regime, while the rest will likely evolve into low-mass stars.

\end{document}